# Stochastic model for Soj relocation dynamics in *Bacillus subtilis*

Konstantin Doubrovinski[*†] & Martin Howard[*‡]

[*]Department of Mathematics, Imperial College London, South Kensington Campus, London SW7 2AZ, UK

[†]Biology Education Centre, Uppsala University, Norbyvägen 14, 752 36 Uppsala, Sweden

[‡]To whom correspondence should be addressed

## Abstract

**The *Bacillus subtilis* Spo0J/Soj proteins, implicated in chromosome segregation and transcriptional regulation, show striking dynamics: Soj undergoes irregular relocations from pole to pole or nucleoid to nucleoid. Here we report on a mathematical model of the Soj dynamics. Our model, which is closely based on the available experimental data, readily generates dynamic Soj relocations. We show that the irregularity of the relocations may be due to the stochastic nature of the underlying Spo0J/Soj interactions and diffusion. We propose explanations for the behavior of several Spo0J/Soj mutants including the "freezing" of the Soj dynamics observed in filamentous cells. Our approach underlines the importance of incorporating stochastic effects when modelling spatiotemporal protein dynamics inside cells.**

# Introduction

Dynamic proteins are vital components of bacterial organization (1-4), and play central roles in accurate cell division (5), plasmid partitioning (6) and developmental regulation (7). The Spo0J/Soj proteins of *Bacillus subtilis* (8,9), implicated in chromosome segregation and transcriptional regulation, form a further important example of dynamic protein localization (1-17). Fluorescence microscopy has shown that Spo0J organizes into compact foci associated with the nucleoid, while Soj undergoes irregular relocations from pole to pole (8) or nucleoid to nucleoid (9). The underlying mechanism for this intriguing dynamic behavior has, however, remained unexplained.

The Soj relocations are reminiscent of the dynamics of the Min proteins in *Escherichia coli* (1-5,10-15), which undergo pole to pole oscillations. These oscillations are exploited by *E. coli* to direct assembly of the essential cell division protein FtsZ into a ring shaped structure precisely at midcell. Importantly, the Min oscillations in *E. coli* are rather regular. This is in contrast to the Soj dynamics in *B. subtilis* which are highly *irregular*. In some cells a patch of Soj protein can jump from nucleoid to nucleoid one or more times over the course of an hour (9), while in other cells it remains on a single nucleoid. Here we propose that these irregularities are due in part to low copy number fluctuations: the relatively low numbers of the Spo0J/Soj proteins in a cell, together with the intrinsic probabilistic nature of their interactions, can lead to large fluctuations in their dynamic behaviour. Hence, fluctuation effects, which are already known to be important in gene regulation (18-20), may also have a profound influence on *spatiotemporal dynamics*, a critical issue that hitherto has received rather little attention (13). We therefore propose that for the Spo0J/Soj system, unlike the Min system, stochasticity is important for capturing the observed irregularity of the spatiotemporal protein dynamics.

In addition to studying this irregularity seen in wild-type cells, we will also examine the behavior of a variety of mutants which display altered Soj dynamics. For example, in filamentous cells the Soj relocations are abolished and instead Soj is restricted to a single nucleoid close to a cell pole. Furthermore, a particular mutant (Spo0J19) causes the Soj relocations to occur much more frequently in normal length cells, and also frees Soj from its polar dependence in filamentous cells. As we will see, understanding the complex behavior of these mutants forms a stringent test of our approach: in particular, nucleoid-polar "shuttling" of Soj will turn out to be an important component of the model dynamics.

Soj/Spo0J are members of the highly conserved ParA/ParB protein family, and possess dual roles in chromosome segregation and transcriptional regulation (8,9,21-26). Spo0J binds to several specific sites close to the origin of DNA replication (22) forming small nucleoprotein complexes. Disruption of *spo0J* is known to have a strong disruptive effect on chromosome segregation/organization in vegetative cells (24,27), and cells with this mutation are also unable to sporulate (24,25). Soj has two distinct roles: firstly, in the presence of Soj, the small complexes of bound Spo0J are condensed into compact foci (9,21,23), although this condensation is apparently not required for chromosome segregation in vegetative cells (24). Secondly, Soj is a transcriptional regulator and can repress the expression of several early stage sporulation genes, an ability which is counteracted by Spo0J (8,24). Hence, when Soj is bound to the nucleoid it organizes Spo0J into compact foci, but when unbound (or perhaps also if prevented from "spreading out" over the entire chromosome by Spo0J) it allows sporulation genes to be activated. Soj relocations are one way in which Spo0J can be organized into foci while still allowing the activation of sporulation genes. However, the origin of the Soj relocation dynamics has so far remained unknown: the object of this paper is to investigate this underlying mechanism.

In order to study the Soj dynamics we will focus on mathematical modelling, an approach which has already been successfully employed in the Min system (10-13,15). This method will allow us to go beyond qualitative pictures and instead quantitatively determine whether particular interlinked interactions of the Spo0J/Soj proteins are able to generate Soj relocations. We will, in particular, develop a minimal mathematical model, thus helping to identify the core ingredients necessary for the Soj dynamics. Despite this simplicity, however, we will still be able to propose an explanation for the dynamics of previously described mutants in which Soj relocations are perturbed, as well as predicting the dynamics of an as yet unstudied mutant.

## Methods

**Numerical calculations**

The partial differential equation model was discretized and numerically integrated using an explicit Euler scheme, with a time step of $1 \times 10^{-5}$ s and lattice spacing of 0.02 μm. For the stochastic model, each lattice site in the simulation represented a dx = 0.02 μm length of the bacterium (e.g. 200 sites for a 4 μm length), and could contain an integer number of protein particles. The time step for the simulation was chosen to be dt = $8 \times 10^{-6}$ s. The simulations were typically run for 2-3 hours of simulated bacterial time. In all cases zero flux boundary conditions were imposed at the cell/nucleoid ends.

## Results

**Deterministic Model**

The components of our model are illustrated in Fig.1, and we begin our analysis by writing these interactions as a deterministic set of nonlinear partial differential equations. We use a one dimensional reaction-diffusion model, which turns out to be sufficient for modelling the Spo0J/Soj dynamics. The equations describe the following interactions of Spo0J/Soj plus diffusion, and are closely based on the available experimental data, although some additional assumptions are also required, as described below: Soj-ATP binds to the nucleoid, preferentially where other Soj is already bound (9,28); Spo0J is then converted to a condensed form (9) at a rate assumed to be proportional to the nucleoid bound Soj density; Soj is then expelled into the cytoplasm (9) as Soj-ADP at a rate assumed to be proportional to the density of condensed Spo0J, where we also assume that bound Soj catalyzes its own disassociation;  Spo0J is then assumed to spontaneously relax back to an uncondensed form; importantly we assume that the expelled Soj cannot immediately rebind to the nucleoid, instead it must first bind to the membrane in the presence of MinD (29), preferentially where other Soj is already bound; finally Soj is assumed to be spontaneously released back into the cytoplasm, from where it can rebind to the nucleoid. Later we will discuss the motivation and evidence for the two cytoplasmic forms of Soj, with one form being able to bind only to the nucleoid while the other can bind only to the polar membrane. Note that we do assume highly simplified Spo0J dynamics: this is a caricature of the highly complex reorganization occurring as Soj brings together and orders the dispersed Spo0J molecules. However, our simplified version appears to be sufficient for capturing the essence of the Soj dynamics. The above processes give the following equations:

$$\frac{\partial [So_1]}{\partial t} = D_1 \frac{\partial^2 [So_1]}{\partial x^2} - k_1 [So_1]\left(1+\sigma_1 [so_n]^2\right) + k_6 [so_m] \quad (1)$$

$$\frac{\partial [so_n]}{\partial t} = D_2 \frac{\partial^2 [so_n]}{\partial x^2} + k_1 [So_1]\left(1+\sigma_1 [so_n]^2\right) - k_2 [so_n][sp]\left(1+\sigma_2 [so_n]\right) \quad (2)$$

$$\frac{\partial [So_2]}{\partial t} = D_1 \frac{\partial^2 [So_2]}{\partial x^2} - k_5 [So_2]\left(1+\sigma_3 [so_m]^2\right) + k_2 [so_n][sp]\left(1+\sigma_2 [so_n]\right) \quad (3)$$

$$\frac{\partial [so_m]}{\partial t} = D_3 \frac{\partial^2 [so_m]}{\partial x^2} + k_5 [So_2]\left(1+\sigma_3 [so_m]^2\right) - k_6 [so_m] \quad (4)$$

$$\frac{\partial [Sp]}{\partial t} = D_4 \frac{\partial^2 [Sp]}{\partial x^2} - k_3 [Sp][so_n] + k_4 [sp] \quad (5)$$

$$\frac{\partial [sp]}{\partial t} = D_5 \frac{\partial^2 [sp]}{\partial x^2} + k_3 [Sp][so_n] - k_4 [sp] \quad (6)$$

Here $[So_1]$ and $[So_2]$ are the densities of the cytoplasmic forms of Soj; $[so_n]$ is the density of Soj on the nucleoid, and $[so_m]$ is the density of membrane bound Soj. $[Sp]$ and $[sp]$ represent the density of uncondensed and condensed Spo0J respectively. The σ terms represent cooperative binding/unbinding of Soj to the nucleoid/membrane. The parameters used were:

$D_1 = 4$ μm$^2$ s$^{-1}$, $D_2 = D_4 = D_5 = 0.006$ μm$^2$ s$^{-1}$, $D_3 = 0$ μm$^2$ s$^{-1}$,

$k_1 = 0.001$ s$^{-1}$, $k_2 = 3.3 \times 10^{-4}$ μm s$^{-1}$, $k_3 = 1 \times 10^{-5}$ μm s$^{-1}$,

$k_4 = 0.02$ s$^{-1}$, $k_5 = 0.0025$ s$^{-1}$, $k_6 = 2.5$ s$^{-1}$,

$\sigma_1 = 0.0025$ μm$^2$, $\sigma_2 = 4.5 \times 10^{-4}$ μm, $\sigma_3 = 10.5$ μm$^2$.

The diffusion constant $D_1$ of Soj in the cytoplasm is similar to that measured experimentally in the *E. coli* cytoplasm, although for an unrelated protein (30). No data is available to fix the other parameters, however, we did check that our model results are robust to variations in these parameters (see supporting information). As suggested by experiment (29), Soj was assumed to bind to the membrane only in the presence of MinD, which is localized at the cell poles. Hence the binding parameter $k_5$ was taken to be nonzero only in the 7.5% of the cell length closest to the cell poles at either end (for both cells of normal length and filamentous cells). Simulated wild type cells were taken to be 4 μm long with 2 nucleoids each of length 1.2 μm, positioned with their ends 0.4 μm from the two cell poles.

Immunoblot measurements have found that there are around 1500 copies of Spo0J in each cell (21); while no figure is currently available for Soj. We assumed a total of 1500 copies of both Spo0J and Soj for wild type cells. Consequently, we used initial conditions where the Soj/Spo0J densities were randomly distributed, but with averages of 375 and 625 $\mu m^{-1}$ in the cytoplasm and on the nucleoids respectively.

For the above parameters, dynamic Soj relocation was sometimes observed, but depended strongly on the initial conditions. For example, when Spo0J was initially entirely in the uncondensed form on both nucleoids, oscillations were subsequently observed. However, if the initial quantity of condensed Spo0J was increased on one of the nucleoids (to 200 or 300 $\mu m^{-1}$), oscillations were no longer seen: instead the Soj assembled on that nucleoid and the dynamics then ceased. But, using initial conditions with Spo0J entirely in the *condensed* form on both nucleoids, regular oscillations were again observed after a transient. Hence, we concluded that the deterministic model with the above parameters did not consistently show oscillatory behavior: the dynamics appeared to depend on the initial conditions, with many initial conditions showing no dynamics at all.

In cases where oscillations are observed, the mechanism for these oscillations lies in the intrinsic Spo0J/Soj interactions which spontaneously generate a dynamic instability that pushes the system towards an oscillatory state (4,31). The dynamic instability requires two key features: first suppose that the concentration of condensed Spo0J on a particular nucleoid is high, implying that most Soj is found elsewhere. As Spo0J spontaneously decondenses, the concentration of condensed Spo0J eventually becomes low, allowing Soj to undergo rapid, cooperative binding to the nucleoid. As the concentration of nucleoid-associated Soj increases it promotes Spo0J condensation. As the quantity of condensed Spo0J increases, it expels nucleoid-associated Soj, a process that is also catalyzed by nucleoid-bound Soj. Soj is then able to diffuse and relocate to the other end of the cell, before undergoing a MinD mediated polar membrane interaction, after which the cycle repeats. Hysteresis, resulting from cooperative binding/unbinding is crucial for the emergence of these oscillations. A second key feature is a high cytoplasmic Soj diffusion constant (of 4 $\mu m^2\ s^{-1}$ similar to those measured in *E. coli* (30)), much larger than the other membrane/nucleoid diffusivities. This is a perfectly natural assumption, since Soj is known to form nucleoprotein filaments, possibly with a helical structure (28). Similar filament formation has been observed for membrane bound MinD (14), and given the similarity between MinD and Soj (both are members of a large family of ATPases and have similar three-dimensional structures (28)), it is possible that membrane-bound Soj will behave similarly. This filamentation will naturally ensure a low diffusion constant for nucleoid/membrane associated Soj. This feature, together with the hysteresis discussed above, helps to promote the generation of a pattern-forming instability.

However, in all cases where oscillations were seen in the deterministic model, the relocations were far too regular, with a well-defined period. We therefore concluded that the deterministic model was unable to give a realistic description of the *irregularity* of the Soj dynamics. We also note that spatiotemporally chaotic behavior, which could in principle also cause irregular Soj dynamics, was not observed.

**Equivalent Stochastic Model**

Since there are only around 1500 copies of Spo0J in each cell (21), we reasoned that low copy number fluctuation effects could be important for the Spo0J/Soj system (13).

We therefore turned to the stochastic analogue of the deterministic model in an attempt to capture the irregularity of the Soj relocations. The partial differential equation model was directly transformed into a discrete particle Monte-Carlo simulation fully incorporating fluctuation effects, where each of the above diffusion/reaction rates was replaced with a diffusion/reaction probability. We used the same kinetic parameters as in the earlier deterministic model (but now representing probabilities rather than transition rates). As an example, each particle of $So_1$ at the i-th site in the cytoplasm (away from the cell ends) could hop to the left or right each with probability $D_1\, dt/(dx)^2$ representing diffusion and, if located next to a nucleoid site occupied with $[so_n]_i$ particles of Soj-ATP, could undergo nucleoid binding with probability $k_1(1+\sigma_1[so_n]_i^2/(dx)^2)dt$, where $D_1$, $k_1$ and $\sigma_1$ all had the same values as above. We again used random initial conditions, with a total of 1500 copies of both Spo0J and Soj in the wild type, where the Spo0J particles were initially randomly distributed in their uncondensed form on the nucleoids, and Soj particles were initially randomly distributed in the cytoplasm in the $[So_1]$ form. For the stochastic model, aside from the total copy numbers of the Soj/Spo0J employed, all our results were now independent of the precise details of the initial conditions.

**Stochastic Model Can Capture Irregularity of Wild Type Relocations**

Simulations of the stochastic model clearly yielded irregular Soj relocations (see Fig. 2), very similar to those observed in experiment. Previously, we found that the *equivalent* deterministic model, which does not incorporate fluctuation effects, did not consistently generate relocations. Hence, we see that low copy number stochasticity is capable of stimulating dynamics that are not always present in the equivalent deterministic model (13,32,33). For the parameters chosen, the fluctuations are central in enabling the Soj to "escape" from one nucleoid to another (see also Fig.4). In the equivalent deterministic model, we found that the Soj cluster was often in a stable state on one of the nucleoids. However, in the equivalent stochastic model, the fluctuations now permit switches between these states, where successive stochastic escape events generate the irregular Soj relocation dynamics. As in the deterministic model, the presence of cooperative Soj binding/unbinding and a high cytoplasmic diffusion constant are again important elements in generating these relocations. Importantly, however, we propose that incorporating stochastic effects is important for a satisfactory explanation of the *irregularity* of the Soj relocations.

The results of Monte-Carlo simulations of wild type cells are shown in Fig. 2(A-D). In some cases irregular Soj relocations were observed (see Fig. 2(A,B)), whereas in other cases (around 15% of simulated cells) Soj remained on a single nucleoid during the simulated time period of about 2 hours (see Fig. 2(C,D)). However, in most simulated cells Soj did relocate, for example in Fig. 2B, where the period for relocation from one nucleoid to another is on the order of 30 min. However, there was clearly a very large variation in the possible periods: more detailed information on the distribution of relocation times can be found in Fig.4. Note that the relocations occur rather abruptly with the proteins dwelling at one end of the cell for far longer than it takes to relocate to the other end. Soj also tends to cluster at the ends of the nucleoid closest to the pole. All the above simulation results are in good agreement with experiment (8,9,29). Membrane bound Soj forms a tight cluster but is spatially offset from the nucleoid region with the highest Soj concentration (see Fig. 5). However, for the model parameters chosen, the quantity of Soj on the membrane is about 20 times less than on the nucleoid (see Fig. 2B).

Close inspection of Fig. 2B reveals the precise step where the stochasticity critically affects the dynamics: when Soj is expelled from the nucleoid/polar membrane (see Fig. 1), it can diffuse and rebind to the polar membrane/nucleoid at either end of the cell. Naturally Soj is more likely to bind to the closest pole/nucleoid, but there is a nonzero probability of escape to the more distant pole/nucleoid. This dynamics can be seen in Fig 2B, where the Soj makes several "attempts" to escape to the other end of the cell before finally succeeding. Notice also that when relocating, the membrane/nucleoid Soj concentrations decrease simultaneously at one cell end, while at the other end they increase simultaneously.

We also examined the dynamics of Spo0J, which tended to switch to its condensed form in regions of the nucleoid occupied by Soj (data not shown). However, capturing more detail of the Spo0J condensation into compact foci would require a significantly more detailed model of the Spo0J/DNA reorganization for which there is currently insufficient data (34).

**Nucleoid-Polar Shuttling**

One of the most important features of our model is that Soj shuttles back and forth from the nucleoid to the polar membrane (see Fig. 1): Soj must first return to the polar membrane before it can rebind to the nucleoid, and similarly, once expelled from the membrane, Soj must first bind to the nucleoid before it can again rebind to the polar membrane. As we will see, this nucleoid-polar shuttling is a possible explanation for the "freezing" of Soj dynamics in filamentous cells, and its absence can explain the freer relocations seen in Spo0J19 mutants.

One previously suggested possibility is that the location of Soj is dictated by the nucleotide to which it is bound (8,29). Recent results have indicated that Soj is an ATPase (28,35) and is present in its ATP form on the nucleoid, where Spo0J acts to stimulate the ATPase activity of Soj (28). One possibility is that membrane associated Soj is in the ADP form. While present at one end of the cell, Soj is shuttling between nucleoid associated ATP and polar membrane associated ADP forms. In this scenario, once expelled from the nucleoid in the ADP form Soj cannot rebind until it undergoes a MinD mediated polar membrane interaction, after which it could be released back into the cytoplasm in its ATP form. However this scenario is made less likely by studies of the SojG12V mutant (28,29), which is able to bind to ATP, although it cannot subsequently form dimers. This form of the protein does nevertheless appear to be able to bind to the polar membrane, forming compact polar bands (29). Furthermore, MinD, which is structurally very similar to Soj, binds cooperatively to the membrane in its ATP form. For these reasons we believe that other scenarios may be more likely. For example, Soj may reacquire ATP in the cytoplasm but still need to undergo a further MinD mediated conformational change before being able to rebind to the nucleoid. This alteration might perhaps permit ATP-dependent Soj dimerization, which is known to be necessary for DNA binding (28). However, these variations all give rise to rather similar mathematical models: for the dynamics to continue, we simply predict that Soj must return to the polar membrane to undergo MinD mediated alteration; the details of this process remain to be fully elucidated. The important nature of this MinD interaction is underlined by experiments: in the absence of MinD, the Soj relocations are impaired, particularly in exponential phase cells where the Soj relocations are virtually abolished (29). In stationary phase cells some limited Soj dynamics are observed (29). However, we have neglected this feature in our model, an assumption which appears to be a reasonable first approximation, particularly so in exponential phase cells.

Despite the continuous shuttling between nucleoid and polar membrane associated forms of Soj in our model, we find that the total numbers present at each location remain roughly constant, while the protein is dwelling at that end of the cell (see Fig. 2B). Hence, we propose that the relocation observed in experiments is between Soj shuttling back and forth between the nucleoid/polar membrane at one end of the cell and a similar shuttling at the other end. However, for the parameters used in our modelling, we find that at any given instant far more molecules of Soj are present on the nucleoid than at the polar membrane (see Fig. 2B).

**Modelling the Spo0J19 Mutant**

Mutagenesis studies have isolated a remarkable Spo0J mutant (Spo0J19) displaying more frequent relocations then in wild type cells (27). We have found two possible modifications of our model that could account for the behaviour of the Spo0J19 mutants.

- Model I

Here we propose that the Spo0J19-expelled Soj-ADP can both rapidly reacquire ATP in the cytoplasm and then relocate to a nearby nucleoid without first undergoing a MinD mediated polar interaction. Possibly the Spo0J19 expelled Soj-ADP has a slightly different conformational form to the Soj expelled by wild type Spo0J, which could account for this difference in behavior. This scenario leads to a simplified set of equations, where Soj is simply cycling between nucleoid bound and cytoplasmic forms, and where we now neglect any possible binding to the cell membrane:

$$\frac{\partial [So]}{\partial t} = D_1 \frac{\partial^2 [So]}{\partial x^2} - k_1 [So]\left(1 + \sigma_1 [so_n]^2\right) + k_2 [so_n][sp]\left(1 + \sigma_2 [so_n]\right) \quad (7)$$

$$\frac{\partial [so_n]}{\partial t} = D_2 \frac{\partial^2 [so_n]}{\partial x^2} + k_1 [So]\left(1 + \sigma_1 [so_n]^2\right) - k_2 [so_n][sp]\left(1 + \sigma_2 [so_n]\right) \quad (8)$$

$$\frac{\partial [Sp]}{\partial t} = D_4 \frac{\partial^2 [Sp]}{\partial x^2} - k_3 [Sp][so_n] + k_4 [sp] \quad (9)$$

$$\frac{\partial [sp]}{\partial t} = D_5 \frac{\partial^2 [sp]}{\partial x^2} + k_3 [Sp][so_n] - k_4 [sp] \quad (10)$$

The parameter values in the above equations are otherwise the same as in the original model. Simulations of this modified model are shown in Fig. 2(E,F). Here, the relocations occur much more rapidly, as one would expect, since the Soj no longer has to undergo an additional polar membrane interaction. Our simulations give an average time for relocation from one nucleoid to another of 10 min, but with variations of around 6 min. The mean period is around 2 times faster than in simulated wild-type cells, and again in good agreement with experiments (27). The assumptions of model I could be directly tested by investigating whether the relocations of Soj in a *spo0J19* background still depend on MinD: model I predicts that the relocations should be MinD independent. In any case, our simulations clearly show that nucleoid-polar shuttling is not a fundamental component for generating dynamic protein relocation, although, as we will see in the next section, it does appear to be important for explaining the freezing of the Soj dynamics in filamentous cells.

- Model II

Here we stick with our original model equations, but now we assume that condensed Spo0J19 has an increased ability to expel Soj. Increasing $k_2$ by a factor of 7 led to the results shown in Fig. 2(G,H). As one would expect given the faster rate at which Soj is expelled, Soj relocates much more rapidly than in simulated wild type cells. However, the simulated relocations, with an average period of about 3 min and variations of about 1 min, were now rather faster than those seen in experiments. The modifications made by this model are, of course, less substantial than in model I, and we believe this scenario may be the more plausible of the two possibilities.

**Frozen Dynamics In Filamentous Mutants**

By depleting FtsZ, an essential cell division protein, long filamentous cells with multiple nucleoids can be produced. In our simulations, filamentous cells were taken to have length 16 μm with 8 nucleoids each of length 1.2 μm regularly distributed along the cell length. Concentration levels were increased 4 fold over cells of normal length, with 6000 copies of Spo0J and Soj present in each simulated filamentous cell. The binding parameter $k_5$ was again taken to be nonzero only in the 7.5% of the cell length closest to the cell poles at either end. This is the same fraction as in normal length cells; however, due to the overall longer length of the simulated cells, this fraction now corresponds to a length scale four times longer than in simulated cells of normal length.

Increasing the cell length (and the number of nucleoids) was observed in experiments to freeze the Soj relocation dynamics (9,29). Using our model to simulate filamentous cells (see Fig. 3(A,B)) reveals strikingly similar frozen dynamics. Furthermore, in our simulations we began with equal amounts of Soj distributed on two nucleoids, the closest and next-closest to one of the cell poles (see Fig. 3(A,B)). In this case the Soj first spread to the nucleoid closest to the cell pole, after which the dynamics ceased, as has been observed experimentally (9).

According to our model, Soj must first return to the cell pole before it can again bind to the nucleoid. If, by chance, Soj escapes to a nucleoid away from the pole, it must still return to the pole before undergoing nucleoid rebinding. Hence, in our model, Soj will always be attracted back to the cell poles. Furthermore, once present at one end of the cell, the Soj is effectively trapped there by the membrane MinD distribution. This is because, in filamentous cells, with a higher copy number of MinD, the MinD membrane distribution substantially overlaps the nucleoids closest to the cell poles. In that case, the Soj continuously shuttles between that nucleoid and the adjacent membrane and is unable to escape by diffusion due to the highly cooperative nature of the Soj nucleoid/membrane binding. For the escape dynamics to be effective the Soj membrane binding needs to be spatially offset from the Soj nucleoid binding, thus giving cytoplasmic Soj the ability to diffuse away rather than immediately bind. This scenario is realized in the shorter simulated cells (see Fig. 5), where the offset can be achieved if the Soj membrane binding is efficient only in regions with the highest MinD concentration, i.e. at the cell poles (36), slightly away from the nucleoids. There is some experimental evidence that this is indeed the case (29). However, in filamentous cells, the higher number of MinD copies means that MinD at a high concentration will cover much more of the membrane and thus overlap the endmost nucleoids much more substantially. In this way nucleoid-pole shuttling effectively traps Soj close to one end of the cell in filamentous cells.

**Active Dynamics in Spo0J19 Filamentous Mutants**

Our models for the Soj dynamics can be further compared to experiments in filamentous Spo0J19 mutants, where the Soj relocates from nucleoid to nucleoid with a higher frequency and with no dependence on the poles (27).

- Model I

In our first model for Spo0J19 cells (see equations (6-10)), Soj is now free to relocate without having to continually return to the poles. Hence, in filamentous Spo0J19 cells, Soj should relocate irregularly between nucleoids along the entire cell length. Our simulations of filamentous Spo0J19 cells using model I (see Fig. 3(C,D))) confirm this and show similar behavior to experiment (27).

- Model II

Simulations of this model in filamentous cells are shown in Fig. 3(E,F). Here we again see Soj relocations independent of the pole. The reason for this stems from the high rate of Soj expulsion by condensed Spo0J19. After expulsion and subsequent MinD mediated interaction at the cell pole, there will still be enough condensed Spo0J19 on the nucleoid closest to the pole to prevent Soj from reforming a cluster there. Hence the Soj will only be able to relocate to a nucleoid further from the pole, as seen in Fig. 3(E,F). The period in this case is again rather short; however, modifying the parameters in an effort to increase the period invariably resulted in the Soj sticking to the nucleoid closest to the cell pole.

**Concentration Levels**

We tested the effect of varying the concentration levels of Spo0J/Soj on both the average period of the relocation and its stochasticity. Here we used a model with otherwise wild type parameters. Increasing the Spo0J and Soj concentrations, while keeping them equal, we found that the period of the relocations decreased. For example, with Soj=Spo0J=3000, we found a period of $5\pm2$ min. However, the relocations were still not totally regular as significant periods emerged where no relocations were observed (see Fig. 6). If the Spo0J and Soj copy numbers were both increased to 5000, the relocations ceased. We also tried increasing the Soj concentration, while keeping the Spo0J copy number fixed at 1500. In that case, for 2000 copies of Soj, we found a period of $5\pm3$ min, but increasing the copy number still further to 3000 or 5000, abolished the relocations. Hence we predict that overexpressing Soj, or Soj/Spo0J together will lead to a reduced period and somewhat reduced fluctuations, but eventually further overexpression will abolish the Soj dynamics. Clearly, from cell to cell, there will be some variation in the copy numbers of the Spo0J/Soj proteins, possibly resulting from unequal partitioning at cell division. Hence some of the variation seen in the Soj dynamics may be coming from the cell to cell variation in the copy numbers rather than intrinsic low copy number stochasticity. However, there is evidence that the Soj dynamics is stochastic even within a single cell (9). Hence, we believe that intrinsic fluctuations resulting from low copy numbers are important for a satisfactory description of the Soj dynamics.

**Varying the MinD Distribution**

It has recently been suggested that MinD may occupy a rather larger fraction of the total cell length than previously believed (37). We therefore varied the length scale over which the Soj membrane binding parameter $k_5$ was taken to be nonzero. However, doubling this length scale from 7.5% to 15%, significantly impaired the Soj relocations in otherwise simulated wild type cells of normal length. Increasing the length scale still further so that Soj could bind anywhere on the membrane (as might be found in a DivIVA mutant, where MinD is uniformly distributed on the membrane), abolished the Soj relocations in our simulations. Hence we predict that MinD overexpression, or the deletion of DivIVA, should have a deleterious effect on the Soj relocations in normal length cells. It would be interesting to test these predictions experimentally. The reason for this behavior in our model is that, for the 15% length scale, the MinD membrane distribution substantially overlaps the closest nucleoids to the poles even in cells of normal length. This then pins the Soj close to the cell poles and abolishes the Soj relocations. For this reason our model predicts that Soj should only bind to regions of the highest MinD concentration, closest to the cell poles. This ensures that, in normal length cells, the membrane and nucleoid Soj populations do not overlap, thus enabling Soj to relocate.

## Discussion

By developing a minimal mathematical model, we have shown that a few simple Spo0J/Soj interactions plus diffusion, together with the key elements of stochasticity and nucleoid-polar shuttling, are able to generate experimentally realistic dynamics. Although spatiotemporal oscillators have now been observed in several cellular contexts (1-9), the Spo0J/Soj system is the first where a stochastic description appears to be important for a proper understanding of the protein relocations. The Spo0J/Soj system thus forms an excellent prototype combining features that will be common to many cellular systems, eukaryotic as well as prokaryotic, where stochasticity is likely to be a vital element in realistic spatiotemporal modelling.

As discussed earlier, the Soj relocations are reminiscent of the Min oscillations in *E. coli*, in that both systems show dynamic relocation. Furthermore both systems consist of an ATPase (MinD/Soj) together with a protein that stimulates ATPase activity (MinE/Spo0J) (28). Nevertheless, there are still striking differences: MinE oscillates in *E. coli* while Spo0J appears to always remain nucleoid bound. Furthermore the irregularity of the Soj relocations is in sharp contrast to the regularity of the Min oscillations. Since the function of the Min dynamics is to provide accurate positional information for cell division positioning, it is clearly necessary for their dynamics to be highly regular. Indeed, the Min protein concentrations must not be too low, or else the positional information they provide will be compromised by low copy number fluctuations (13). By contrast, the Soj dynamics is highly irregular, since in many cells

Soj does not relocate at all, or does so only very erratically. Hence any positional information provided by Soj is likely to be rather inaccurate. For this reason it is difficult to see how Spo0J/Soj can play a vital role in specifying accurate positional information. Clearly, more work is needed to understand why *B. subtilis* has made the Soj relocation dynamics so unreliable.

We thank J. Errington, M.E. Fisher, J. Löwe, J. Stark and P.R. ten Wolde for discussions, and also the referees for valuable comments. M.H. acknowledges support from The Royal Society.

**Supporting Information**

**Model Robustness**

In order to test the robustness of our results, we varied both the kinetic parameters in our model, as well as altering the structure of the equations themselves. We then resimulated the resulting Spo0J/Soj dynamics. We found that our results were substantially insensitive to many, though not all, of the perturbations:

- We tested the effects of varying the membrane/nucleoid diffusion constants: in the wild type model $D_3$ was set to zero. However, setting it equal to 0.006 $\mu m^2\ s^{-1}$ (the same as the other membrane/nucleoid diffusion constants) did not appreciably alter our results (with the exception of widening the Soj membrane distribution, see Fig. 5). Setting $D_4 = D_5 = 0\ \mu m^2\ s^{-1}$ also did not significantly alter the model dynamics. However, setting $D_2 = 0\ \mu m^2\ s^{-1}$ did destroy the Soj relocations. The key point here is that the membrane/nucleoid diffusion constants must be small compared to the cytoplasmic diffusion constants; with the exception of removing Soj nucleoid diffusion altogether, their precise values appear to be unimportant. Varying the cytoplasmic diffusion of Soj also had only a weak effect on the Soj dynamics: any diffusion constant larger than about $D_1=1$ μm$^2$ s$^{-1}$ yielded significant Soj relocations.

- We also varied the reaction rates: Soj relocations persisted even when $k_1$, $k_2$, $k_3$, $k_5$, $\sigma_1$, $\sigma_2$ were each separately increased by a factor of 2. However, increasing $k_4$, $k_6$ by a factor of 2 did abolish the relocations. Separately reducing $k_4$, $k_6$, $\sigma_3$ by a factor of two again led to Soj relocations. Hence, the Soj relocations are fairly robust to changes in the kinetic constants. The variations in $k_1$, $k_2$, $k_3$, $k_4$, $\sigma_1$, $\sigma_2$ were performed on the Spo0J19 mutant model I and the variations in $k_5$, $k_6$, $\sigma_3$ on the wild type model.

- We altered the exponents governing the cooperativity of the Soj binding/unbinding dynamics. Setting either of $\sigma_1$, $\sigma_3$ to zero, with all other parameters having their wild type values, suppressed the Soj relocations. Setting $\sigma_2$ to zero abolished relocations in simulations of the Spo0J19 mutant (both models). Cooperativity therefore appears to be an important element in generating realistic Soj dynamics. However, provided the σ terms are nonzero, there is still flexibility in the choice of the exponents in the cooperativity terms (previously set equal to two for the $\sigma_1$, $\sigma_3$ terms, and unity for the $\sigma_2$ term). For example, using otherwise wild type parameters, but setting the exponent to 1.5 for the $\sigma_1$, $\sigma_2$ terms, and with $\sigma_1$ and $\sigma_2$ increased by factors of 10 and 40, also yielded Soj relocations albeit of an increased frequency and reduced amplitude. Using an exponent of 2.5 for the $\sigma_1$, $\sigma_2$ terms, and with $\sigma_1$ and $\sigma_2$ reduced by factors of 20 and 10, yielded relocations similar to those in the wild type model.

- We also tested several variations on the nature of the cooperative Soj unbinding from the nucleoid. For example, we altered the model so that the Spo0J condensation process was cooperative, meaning that the Spo0J condensed preferentially where condensed Spo0J was already present. A high local density of condensed Spo0J could then be responsible for the cooperativity in the Soj expulsion process. However, despite some effort, these modifications suppressed the Soj dynamics. Hence we currently favor our model where the Soj catalyzes its own disassociation in the presence of condensed Spo0J. However, given the complexity of the Spo0J/chromosome condensation we cannot rule out the possibility that this failure is a consequence of an over-simplification in our modelling. To test this point further more data is needed so that more realistic models can be constructed of the Spo0J/chromosome reorganization dynamics.

- Adding protein production and degradation with a Soj half-life of 30 minutes did not significantly alter the dynamics in simulated wild type cells.
- We also tested a model where the requirement that Soj first bind to the polar membrane before rebinding to the nucleoid was relaxed. In this implementation there was only a single species of cytoplasmic Soj. This form was assumed able to bind to either the nucleoid or MinD/polar membrane (using the same $k_1$, $k_5$ terms as in the wild type model equations): these regions simply competed for Soj binding. However, using this modified model, we were unable to reproduce the results found in experiments. The difficulty here is that in order to obtain relocations in simulated wild type cells, the Soj must be able to bind relatively easily to any given nucleoid. However, in filamentous cells, this has the consequence that it is easy for the Soj to diffuse away and populate nucleoids well away from polar regions. In essence, the presence of extra Soj binding sites close to the poles is insufficient to "pin" the Soj close to the pole and prevent the Soj from diffusing away. Finally, we mention that this model is also made less likely by experimental results in *minD* mutants, where membrane Soj binding is prevented and the Soj dynamics virtually abolished (at least in exponential phase). This behavior is not what one would expect if the MinD/polar membrane were simply competing for Soj binding.

**Fig. 1.** Schematic representation of the model for the Spo0J/Soj dynamics. Soj-ATP undergoes spontaneous cooperative nucleoid binding (9,28); nucleoid bound Soj induces Spo0J condensation (9). Condensed Spo0J stimulates the ATPase activity of Soj, leading to cooperative expulsion (9) of Soj-ADP from the nucleoid into the cytoplasm. Once most of the Soj has been expelled the Spo0J begins to revert back to its uncondensed form. Expelled Soj binds to the cell membrane in the presence of MinD (29). MinD is a polar membrane associated protein meaning that, in our model, Soj can only bind close to the cell poles. Soj is then released back into the cytoplasm where it is now able to rebind to the nucleoid, and the cycle repeats. While cytoplasmic, Soj may diffuse and then rebind to the closest nucleoid/polar membrane or, with a lower probability, it may relocate to the other nucleoid/polar membrane. This leads to spontaneous dynamic Soj relocations.

**Fig. 2.** Stochastic simulations of normal length *B. subtilis* cells. (A,C,E,G) Space-time plots of nucleoid bound Soj concentration; bright colors representing high concentration on a nucleoid. (B,D,F,H) Number of Soj proteins on each nucleoid (+,o), and at the left hand polar membrane (B only, full line, copy number exaggerated by a factor of 10), as a function of time. (A,B) Example of wild type cell simulation showing stochastic relocation. (C,D) Example of wild type cell simulation where the Soj patch fails to relocate to another nucleoid. (E,F,G,H) Spo0J19 mutant simulations; (E,F) model I; (G,H) model II, showing much more rapid Soj relocation dynamics.

**Fig. 3.** (A,C,E) Space-time plots of nucleoid bound Soj concentration; bright colors representing high concentration on a nucleoid. (B,D,F) Number of Soj proteins on each nucleoid as a function of time. (A,B) Stochastic simulations of filamentous cells depleted for FtsZ. (C,D,E,F) Simulated Spo0J19 mutants lacking FtsZ; (C,D) model I; (E,F) model II. In both cases the dependence on the poles is lost and Soj is able to relocate from nucleoid to nucleoid well away from polar regions.

**Fig. 4**. Histogram of the Soj relocation period, for wild type simulations where Soj was seen to relocate at least once over a simulated period of 100 minutes. Note the absence of any peak, reminiscent of a characteristic period. This is consistent with our analysis of the Soj relocations as being noise driven "escape" events.

**Fig. 5.** (A) Space-time plots of membrane bound Soj, and (B) both membrane and nucleoid bound Soj, for the case where $D_3=0.006$ μm$^2$ s$^{-1}$, with otherwise wild type parameters; bright colors represent high concentrations on a nucleoid or polar membrane. Notice that the membrane bound Soj forms tight clusters close to the cell poles, and that these clusters are offset from the nucleoid bound Soj clusters.

**Fig. 6**. Typical data for the number of Soj proteins on each nucleoid as a function of time, from simulated cell with copy numbers of Soj and Spo0J equal to 3000 but with otherwise wild type parameters. Note the rather regular dynamics, punctuated by periods with far more random dynamics.

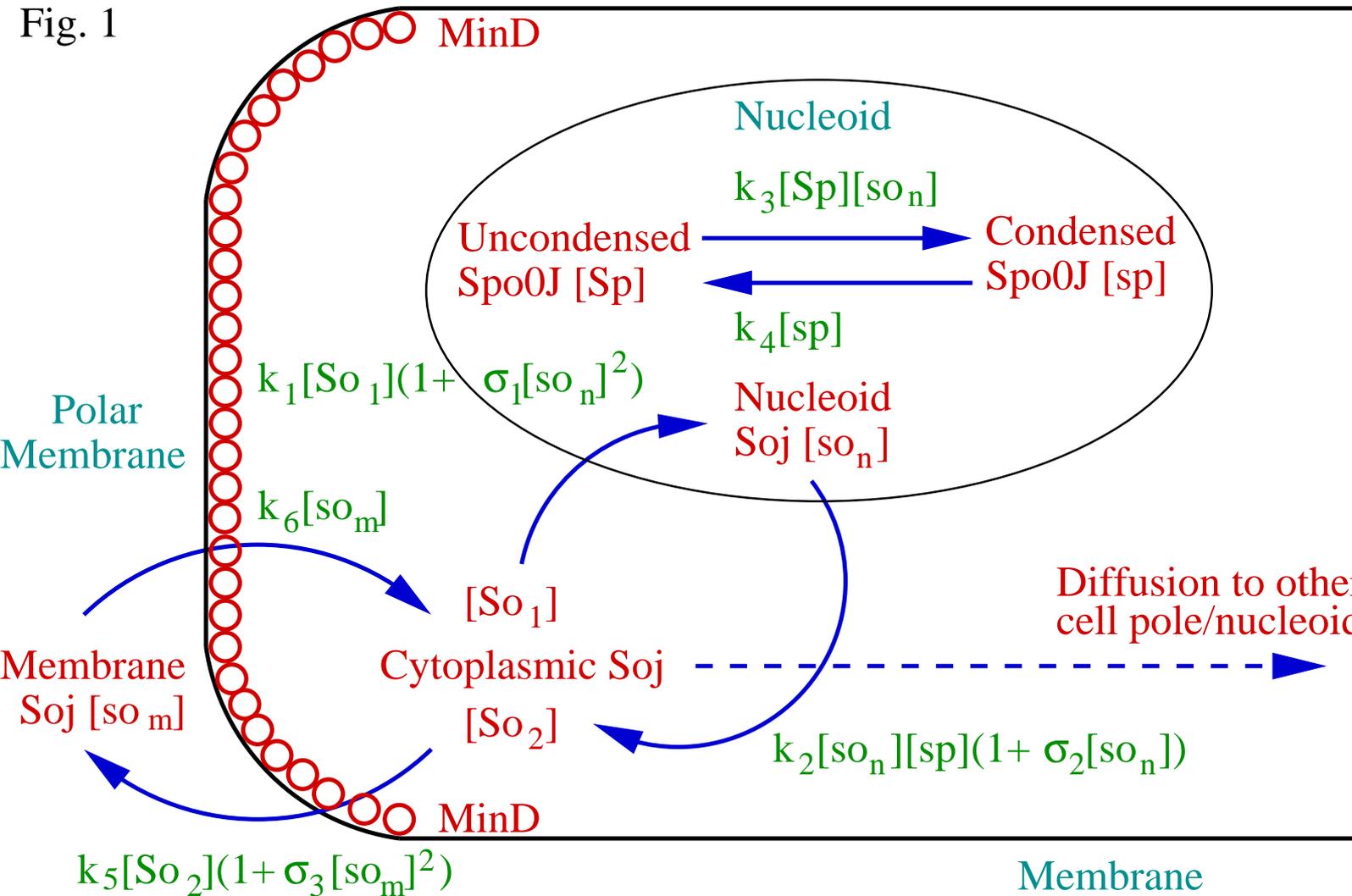

Fig. 1

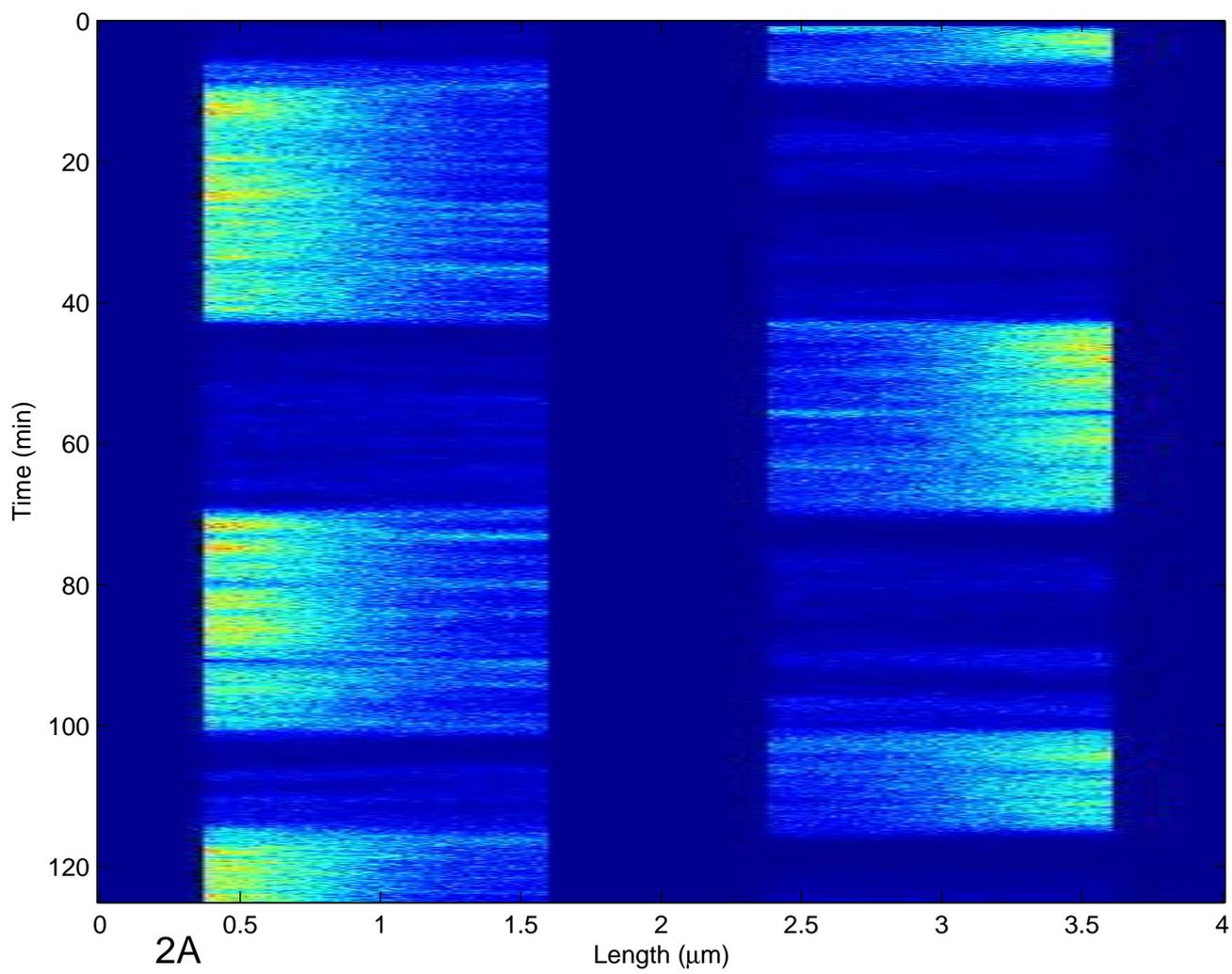

2A

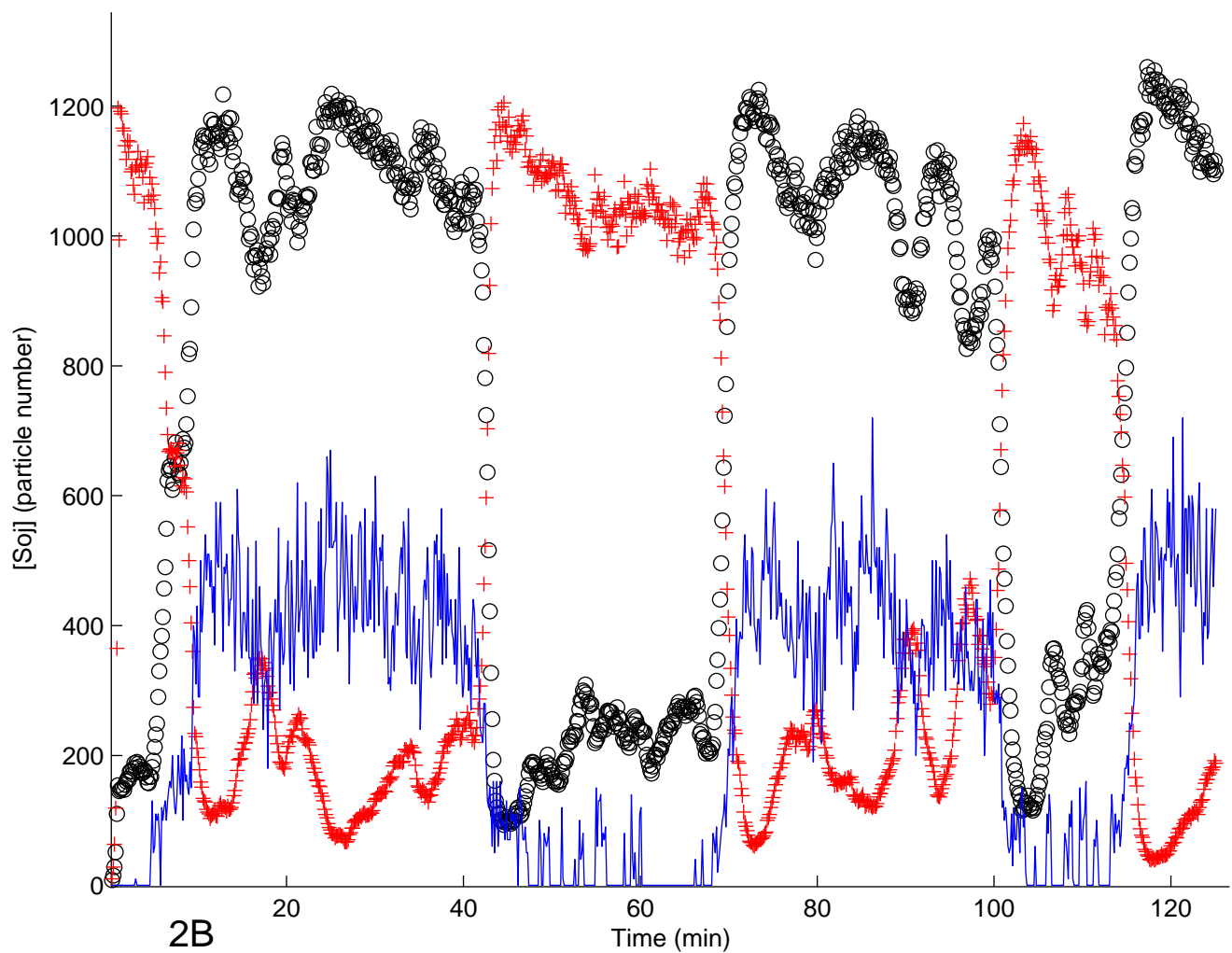

2B

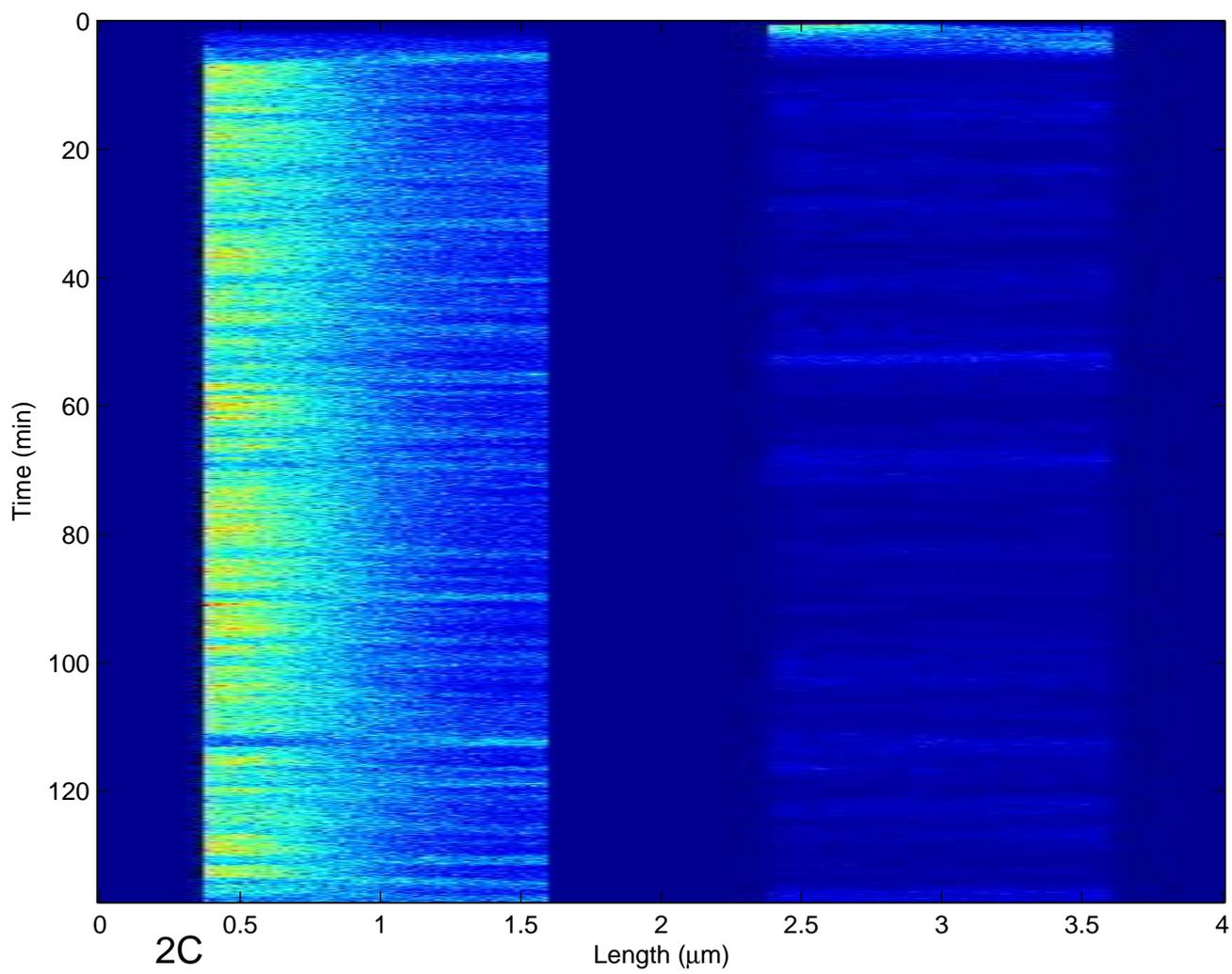

2C

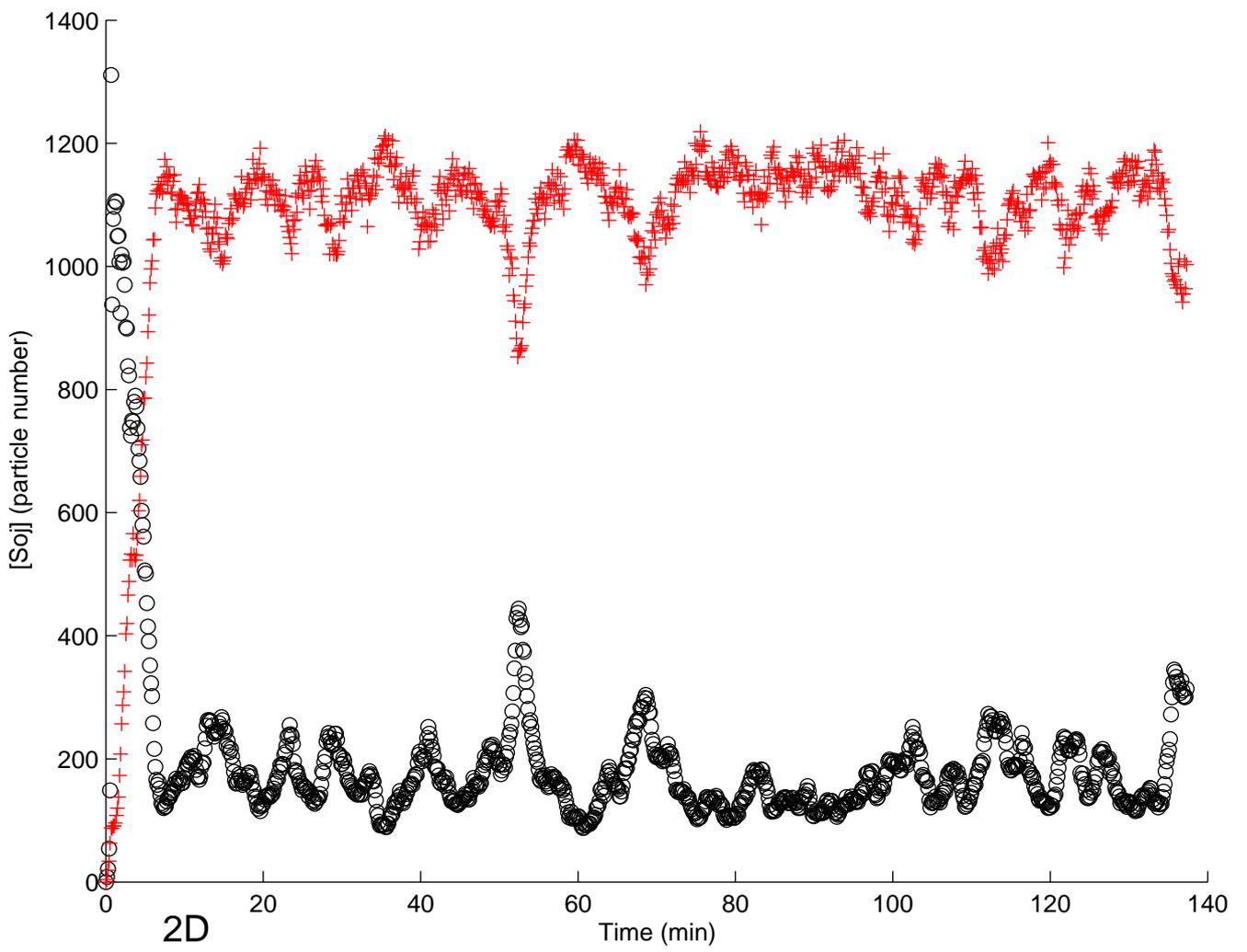
2D

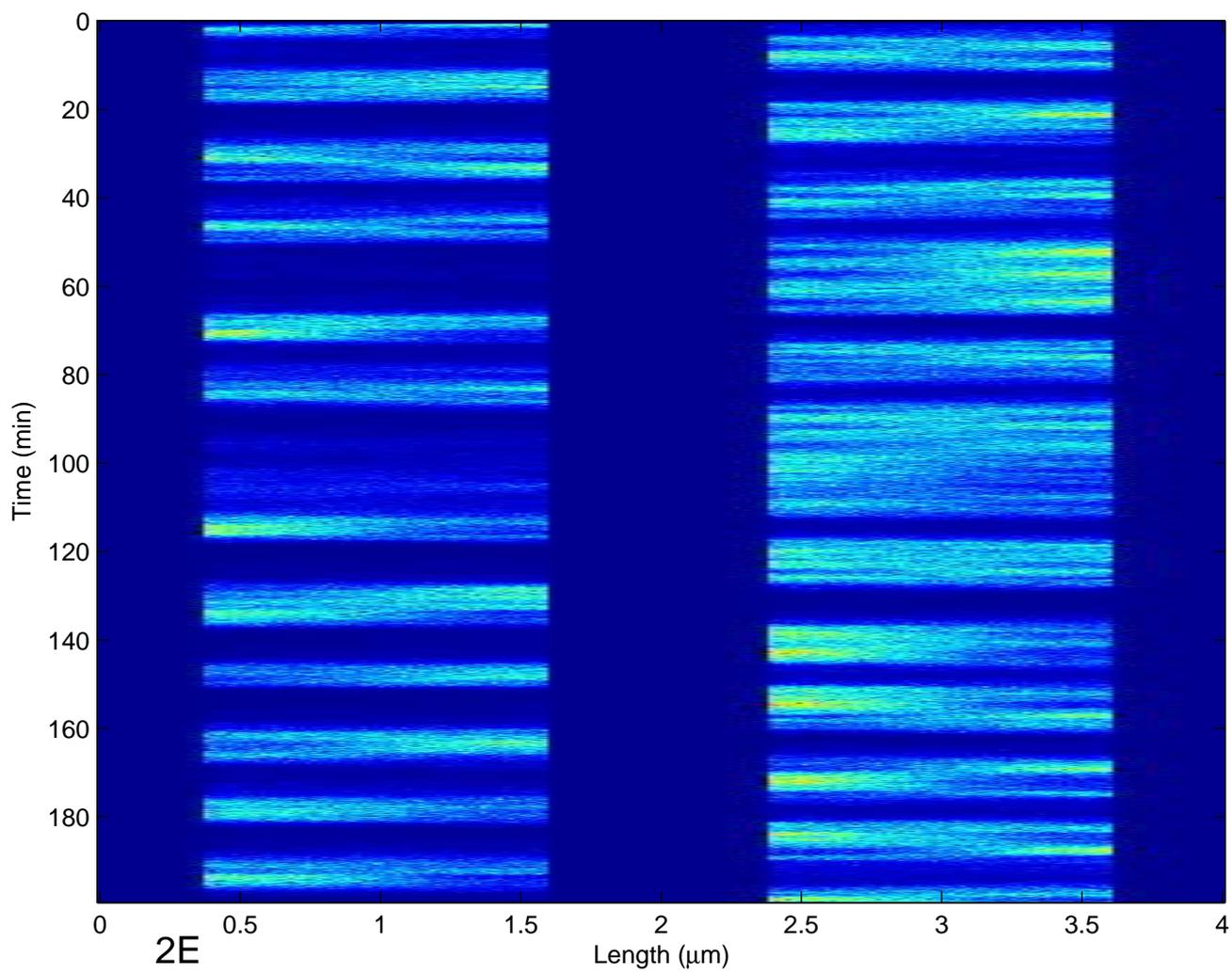

2E

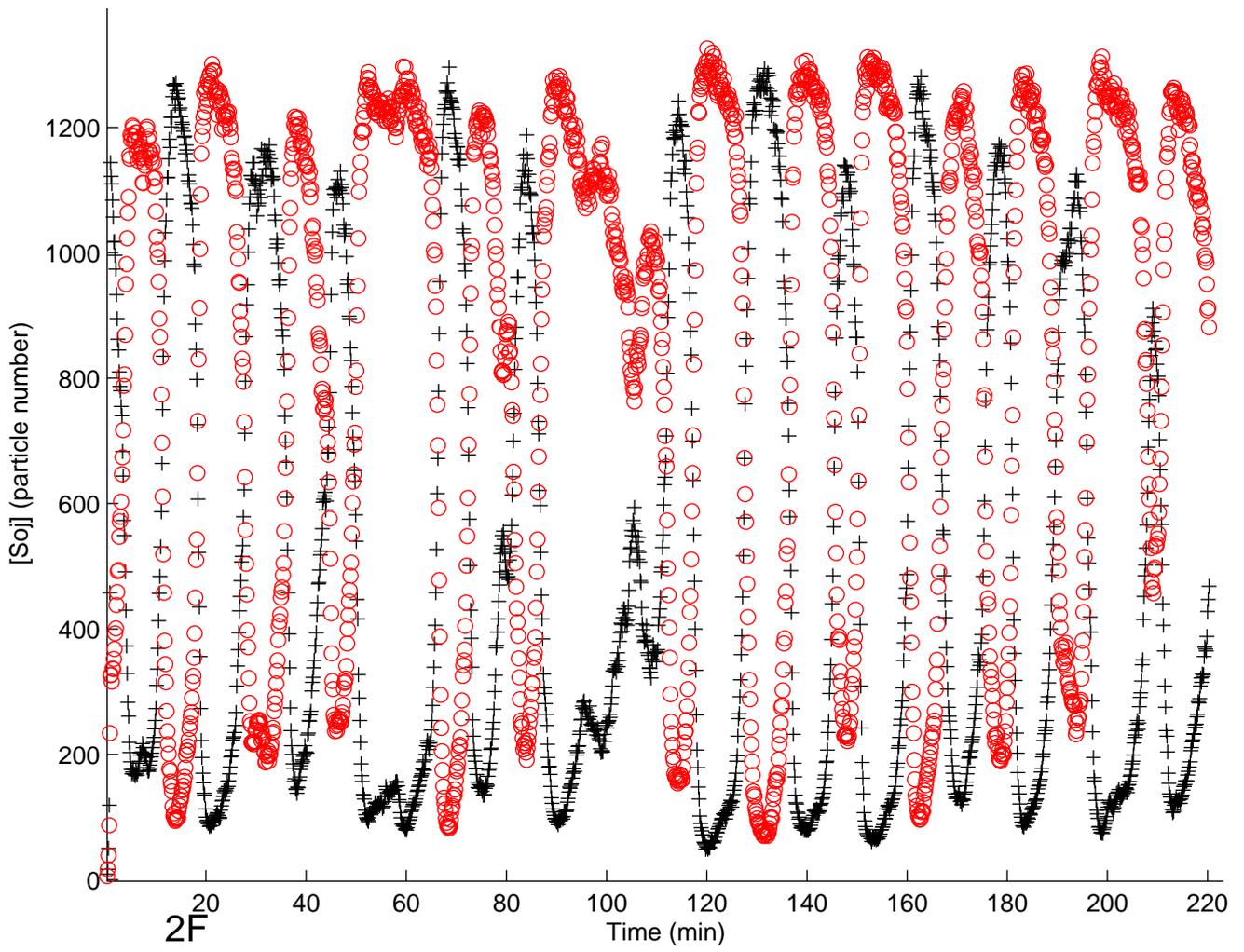

2F

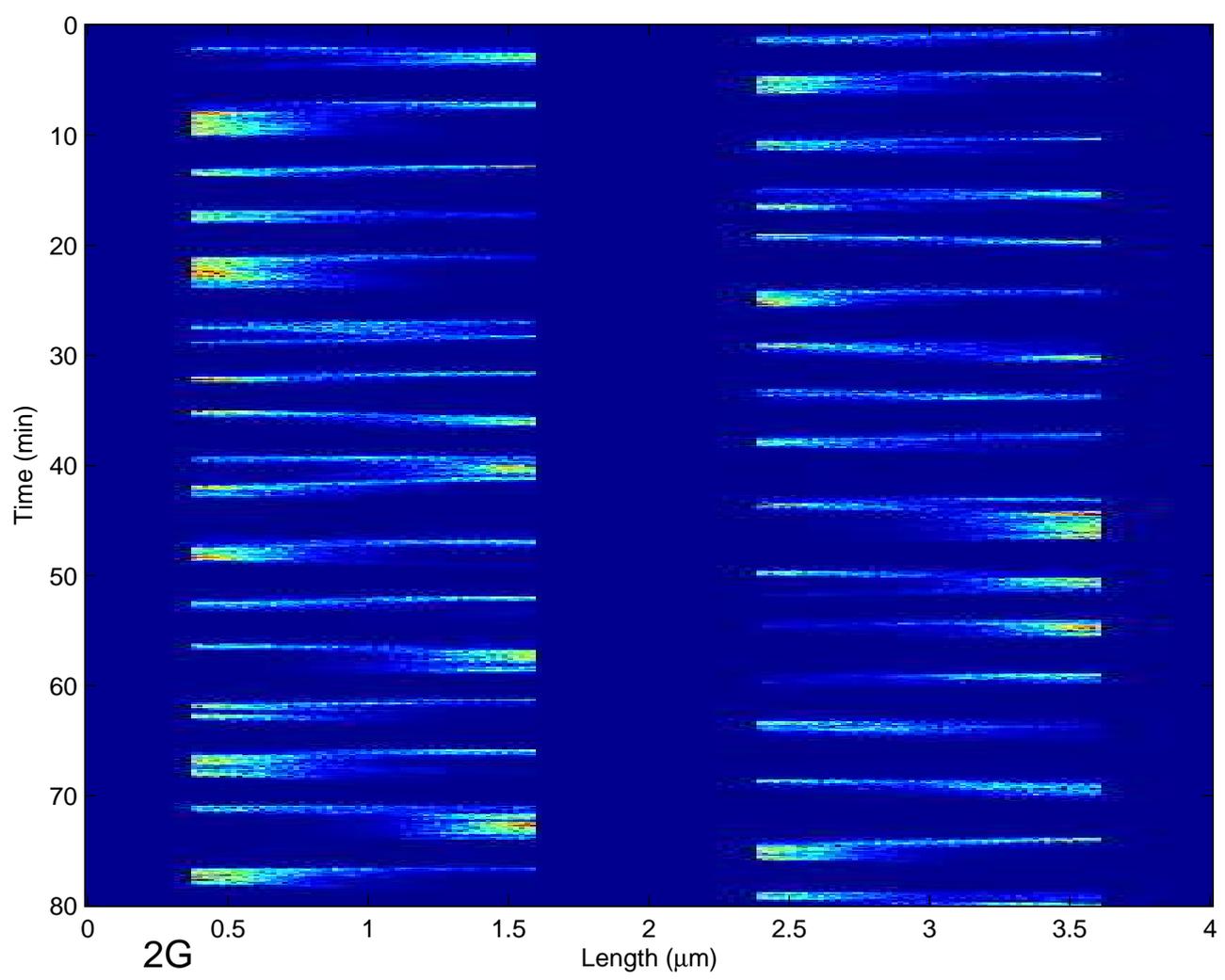
2G

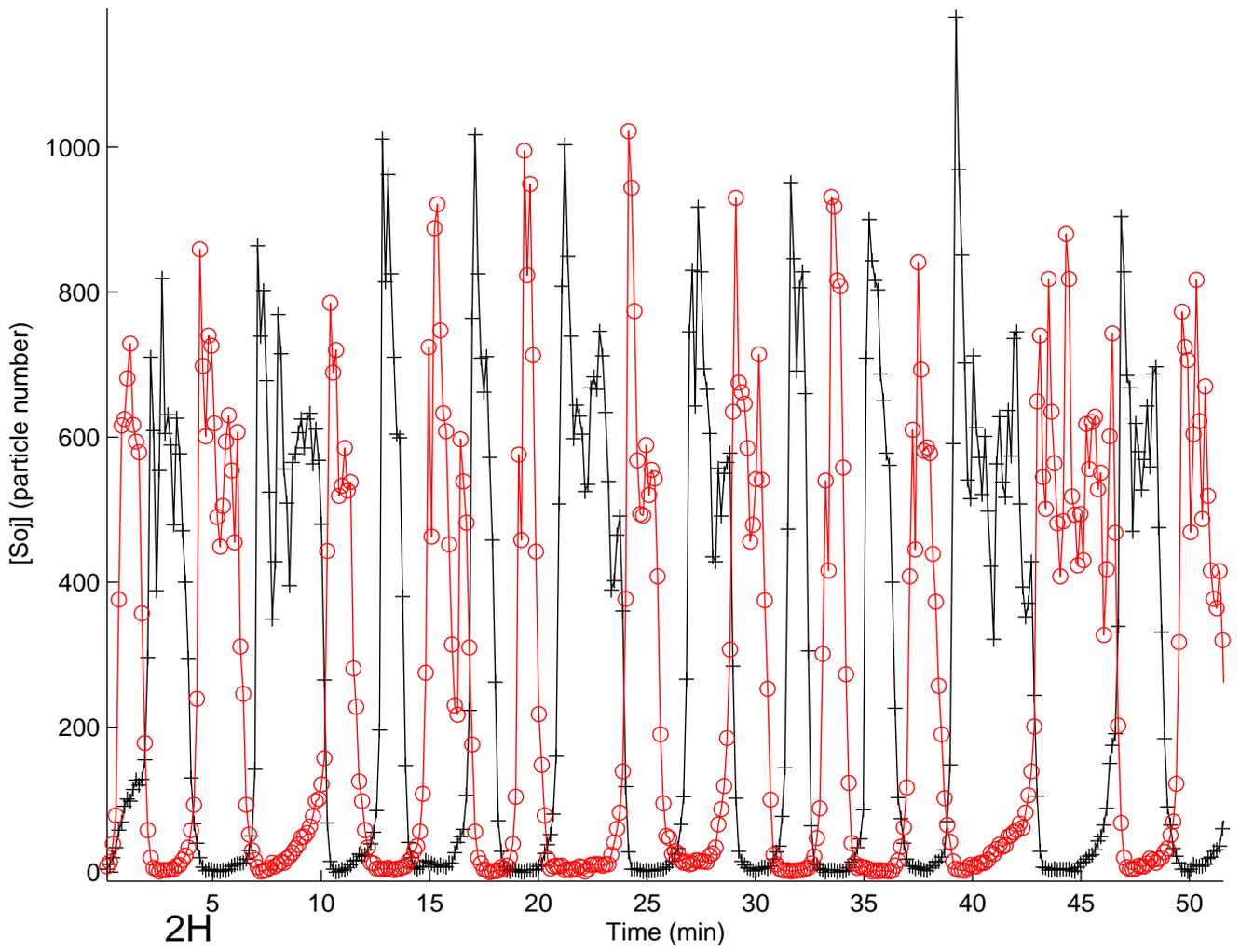
2H

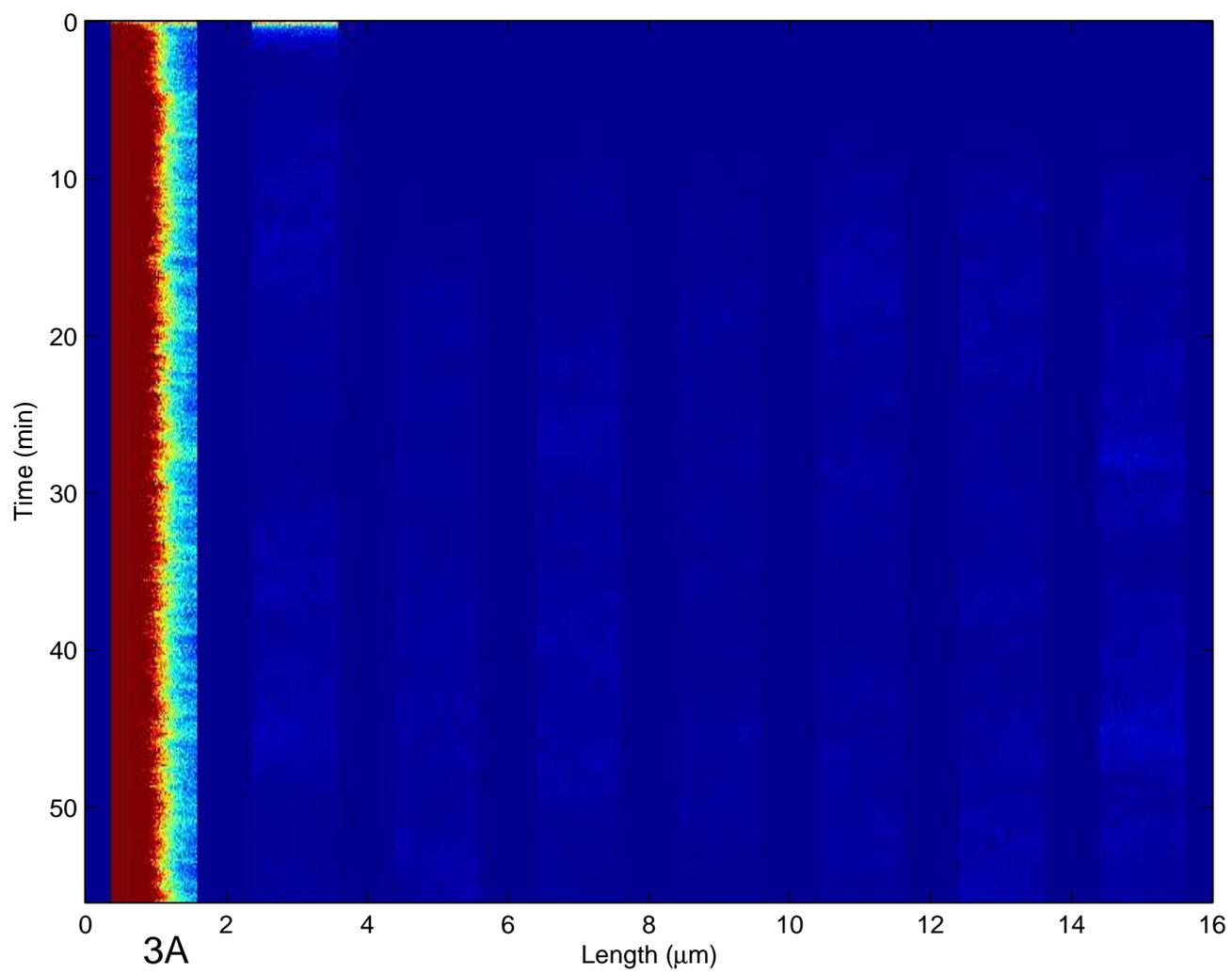

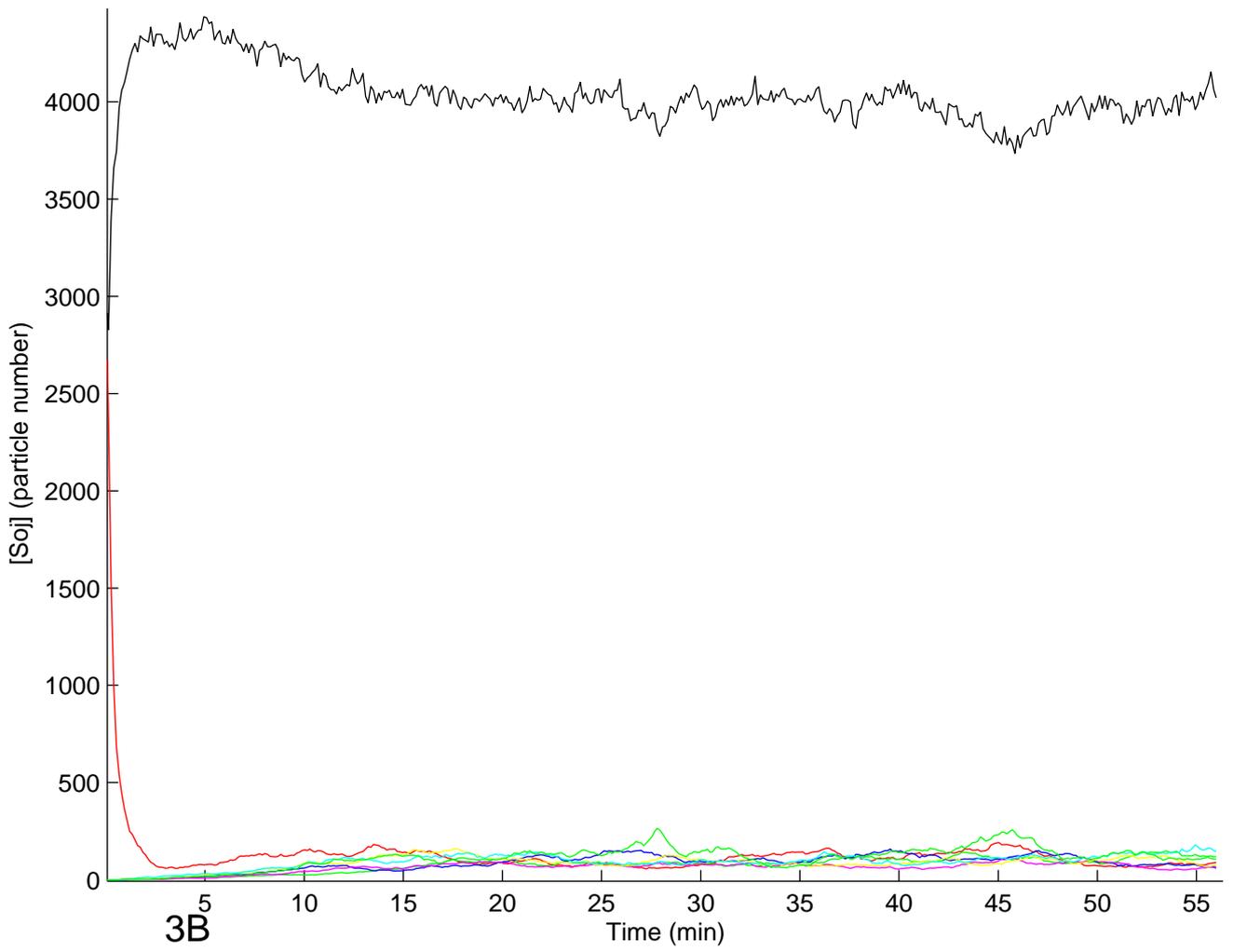

3B

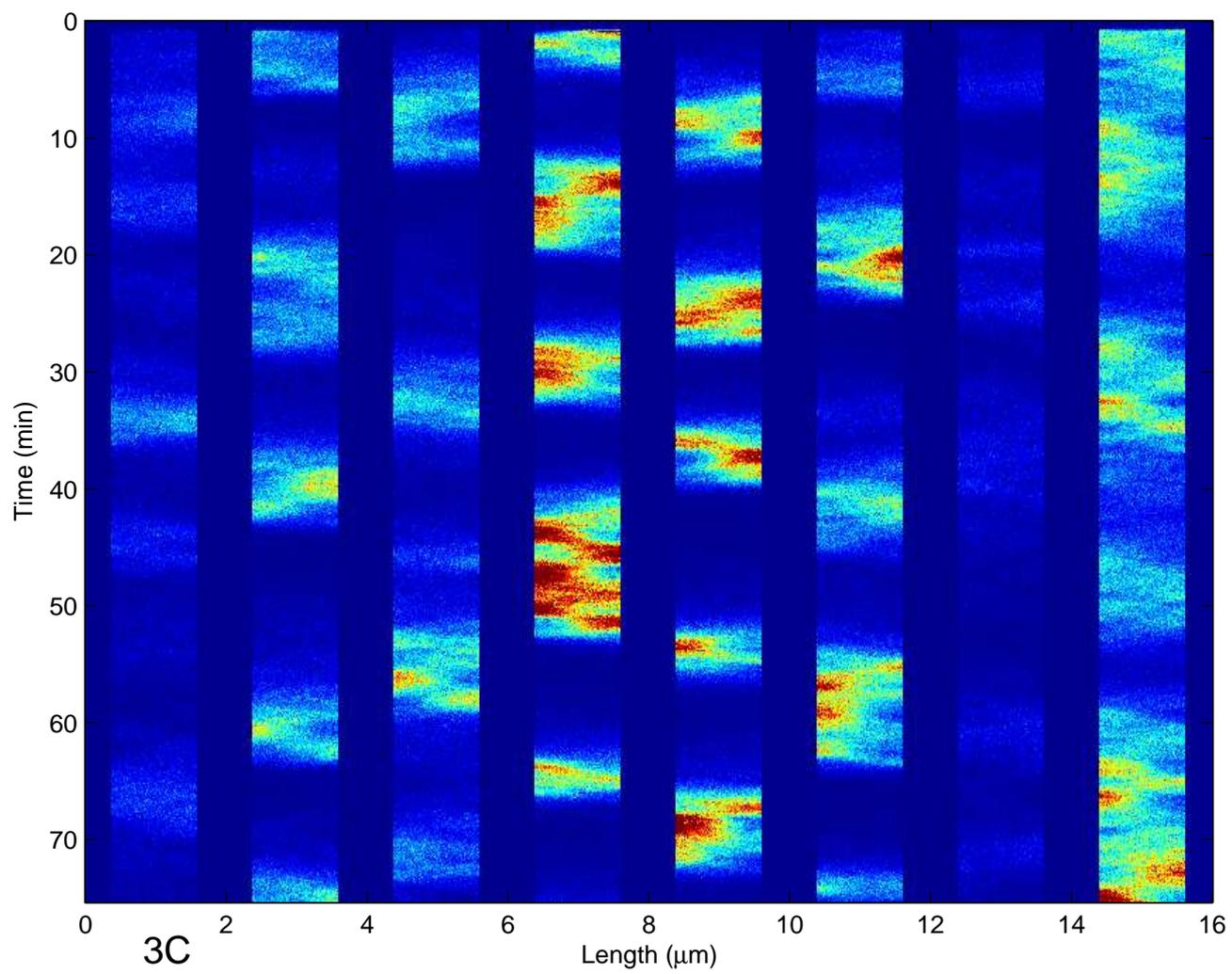

3C

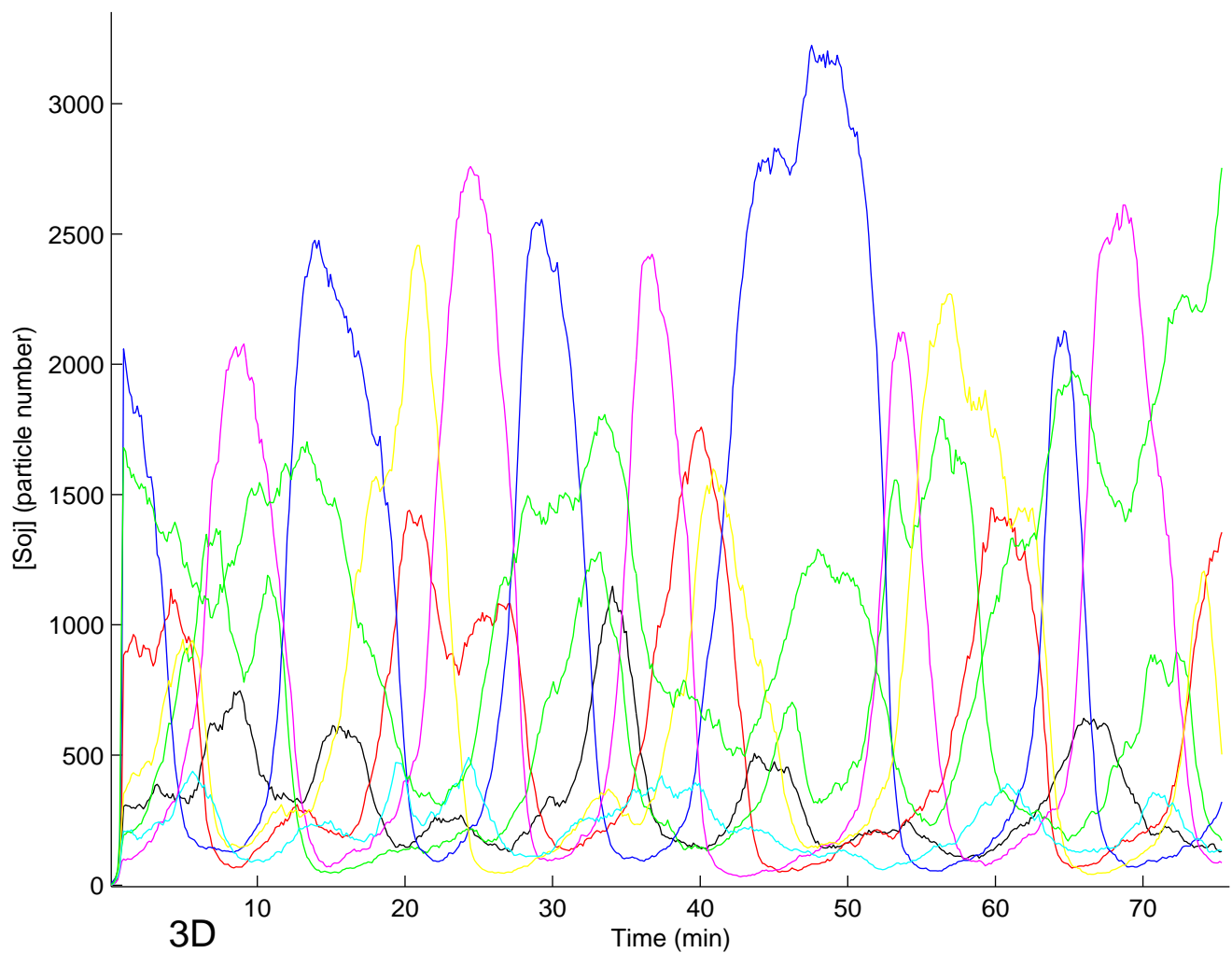

3D

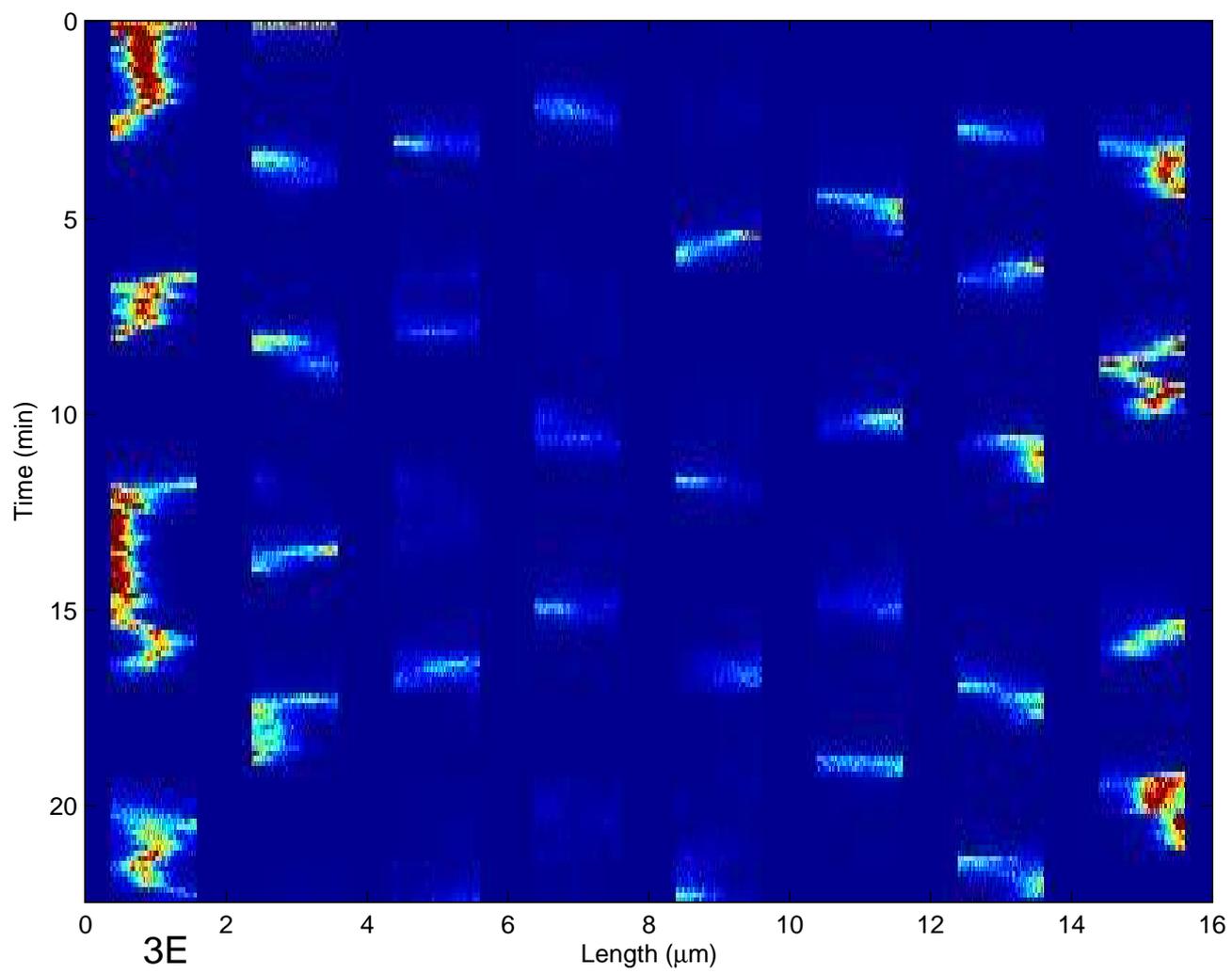

3E

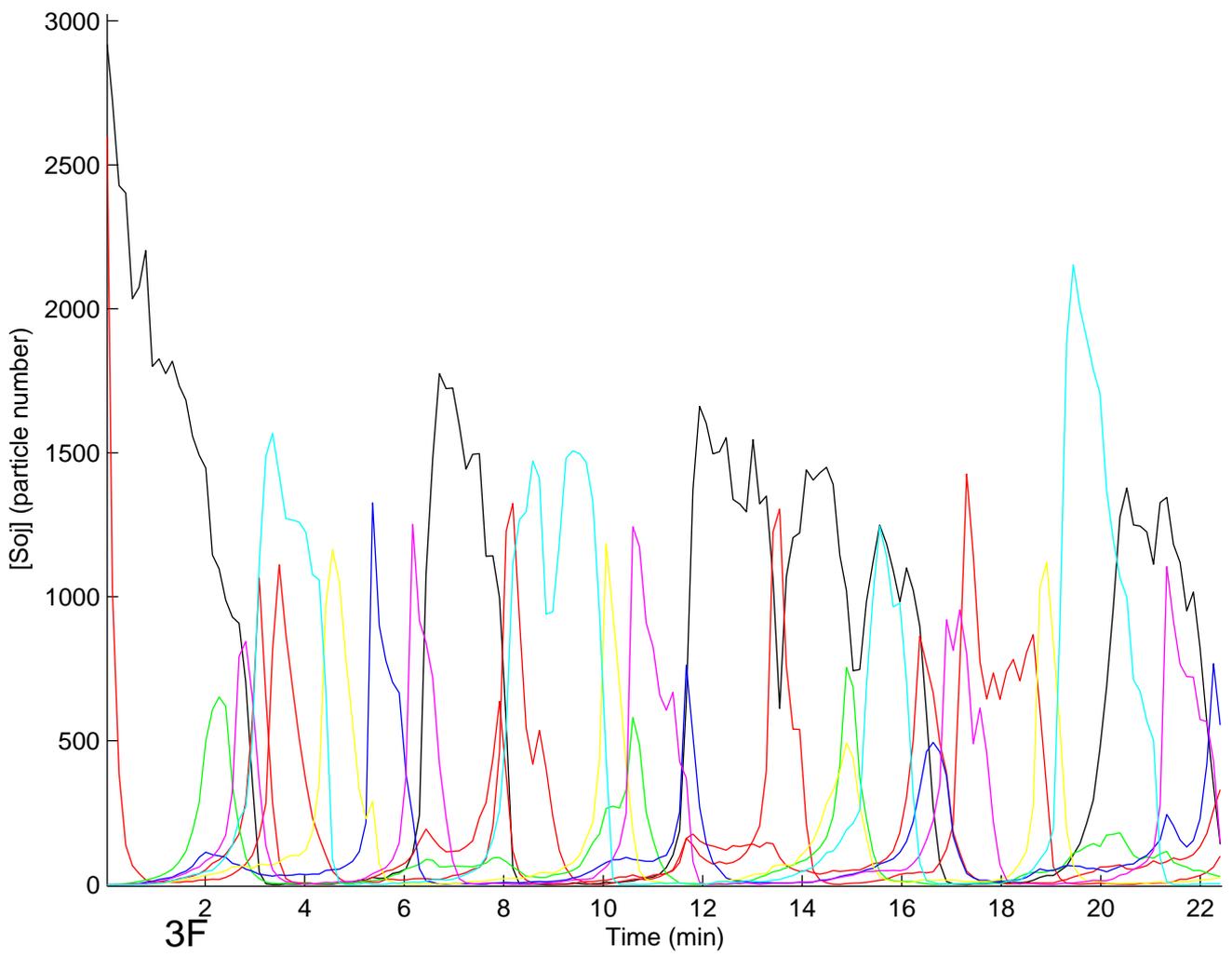

3F

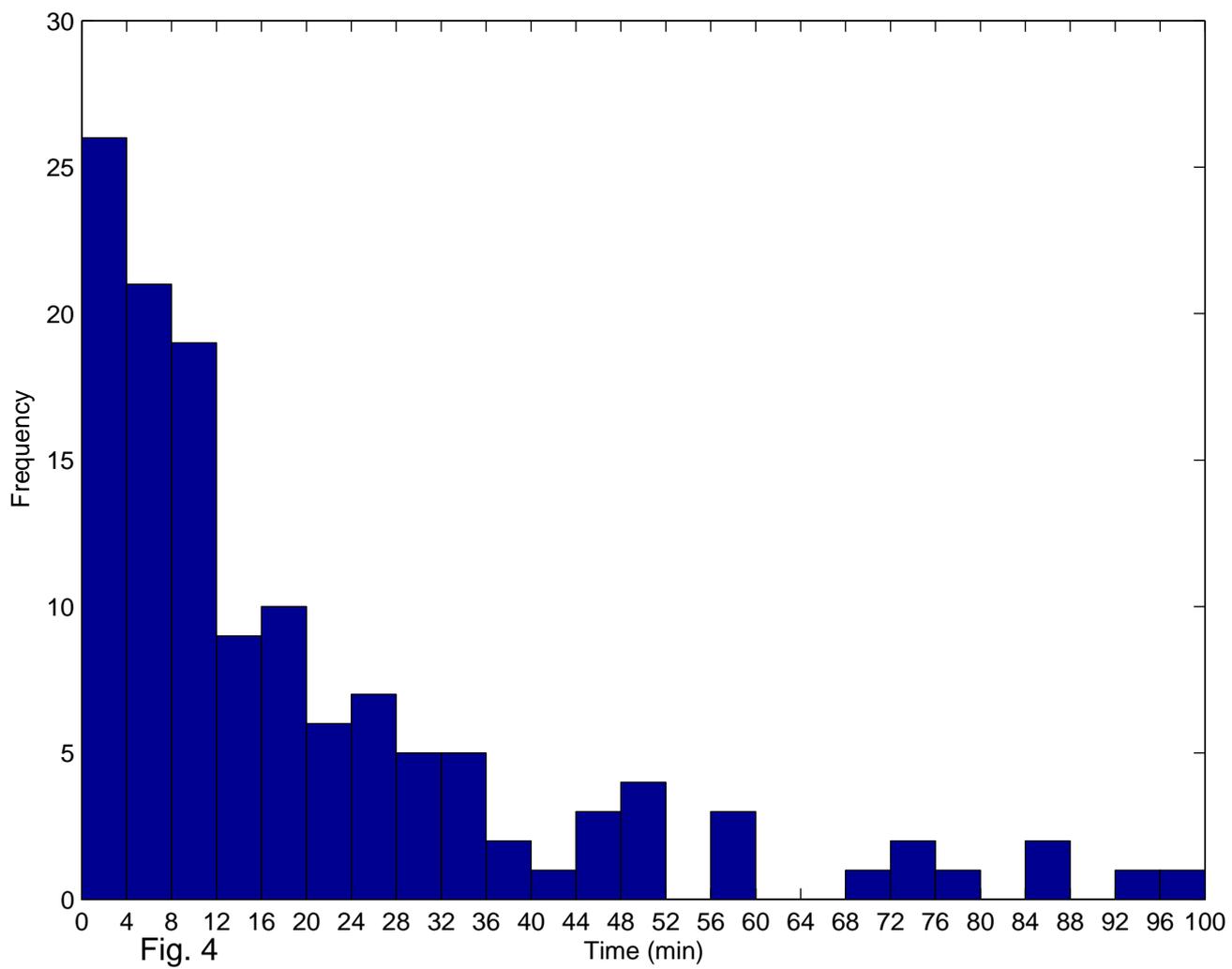
Fig. 4

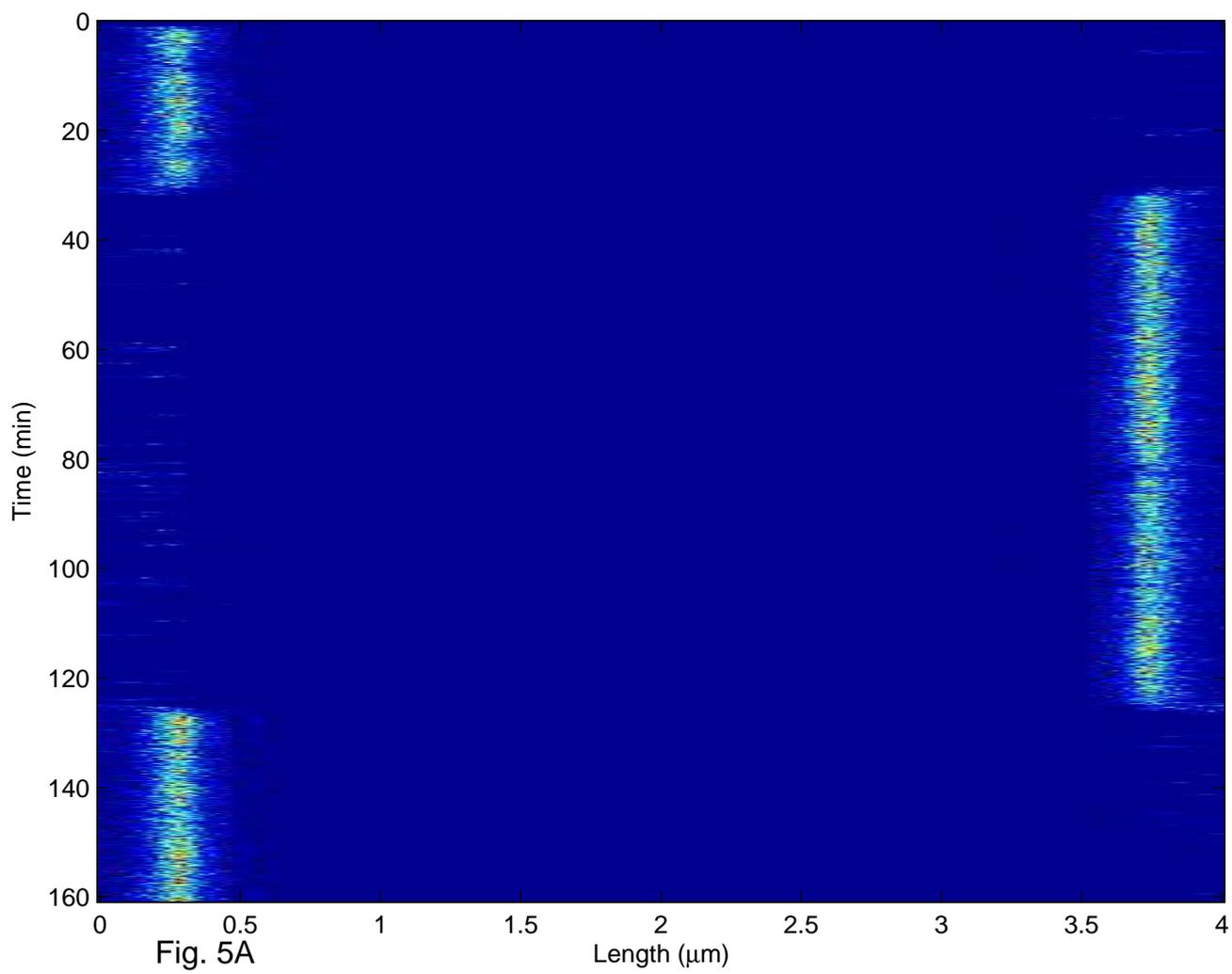
Fig. 5A

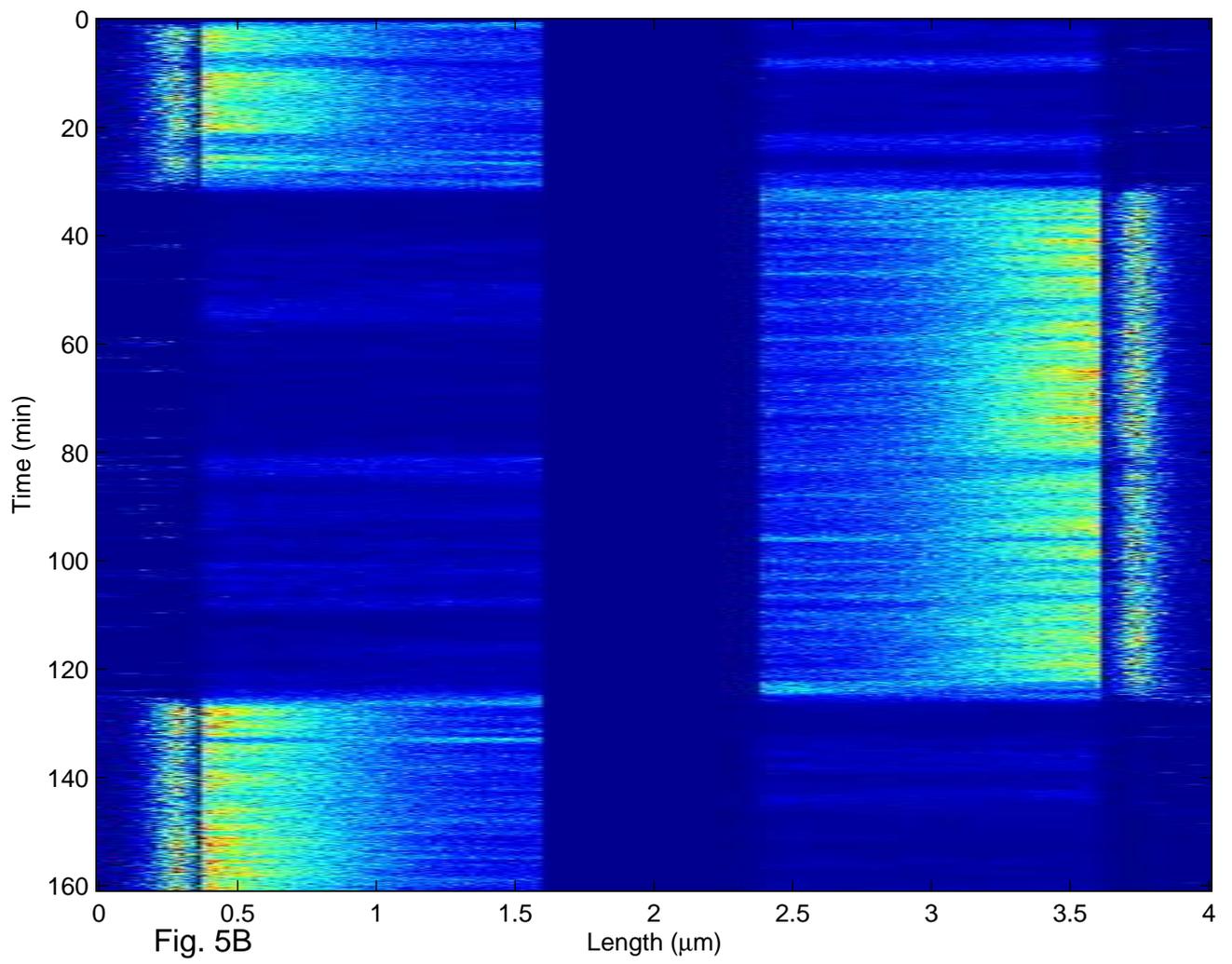

Fig. 5B

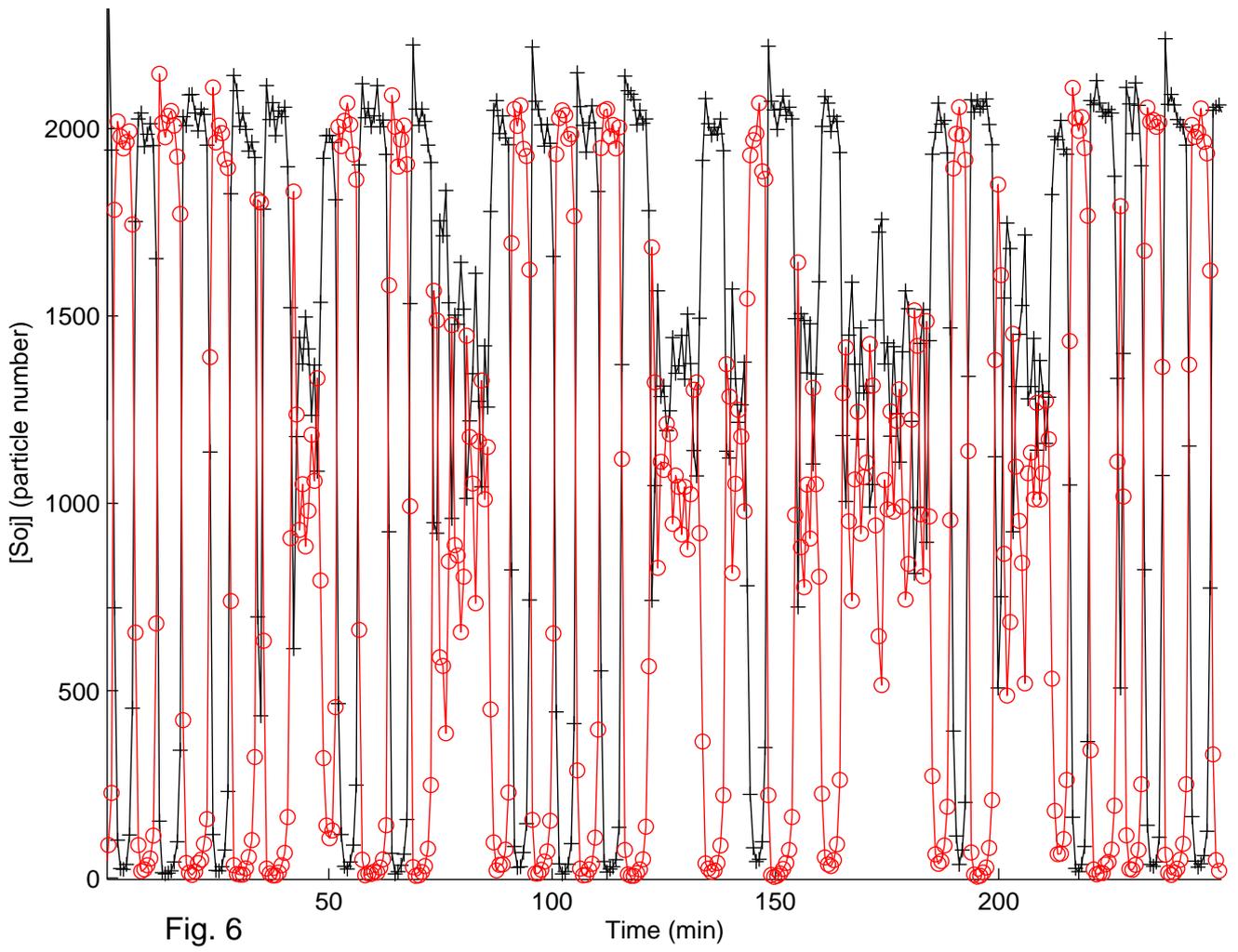

Fig. 6